\documentclass[aps,twocolumn,showkeywords,floatfix,superscriptaddress]{revtex4-2}
\usepackage{graphicx}
\usepackage{hyperref}
\usepackage{amsfonts}
\usepackage{amsmath,bm,amssymb}
\usepackage{mathtools}
\usepackage{comment}
\usepackage{xr-hyper}
\usepackage{hyperref}
\usepackage{placeins}

\begin{document}


\title{Spin-polarized zero-bias peak from a single magnetic impurity at an $s$-wave superconductor: first-principles study}
\author{Kyungwha Park}
\email{Corresponding author: kyungwha@vt.edu}
\affiliation{Department of Physics, Virginia Tech, Blacksburg, Virginia 24061, USA}
\author{Bendeguz Nyari}
\affiliation{Department of Theoretical Physics, Budapest University of Technology and Economics, P.O. Box 91, H-1521 Budapest, Hungary}
\author{Andras Laszloffy}
\affiliation{Department of Theoretical Physics, Budapest University of Technology and Economics, P.O. Box 91, H-1521 Budapest, Hungary}
\affiliation{MTA-BME Condensed Matter Research Group, Budapest University of Technology and Economics, H-1111 Budapest, Hungary}
\author{Laszlo Szunyogh}
\affiliation{Department of Theoretical Physics, Budapest University of Technology and Economics, P.O. Box 91, H-1521 Budapest, Hungary}
\affiliation{MTA-BME Condensed Matter Research Group, Budapest University of Technology and Economics, H-1111 Budapest, Hungary}
\author{Balazs Ujfalussy}
\email{ujfalussy.balazs@wigner.mta.hu}
\affiliation{Wigner Research Centre for Physics, Budapest, P.O. Box 49 H-1525, Hungary}

\date{\today}


\externaldocument[S-]{SI_PRB}
%
\begin{abstract}
Magnetic impurities at surfaces of superconductors can induce bound states referred to as Yu-Shiba-Rusinov (YSR) states within superconducting gaps. Understanding of YSR states with spin-orbit coupling (SOC) plays a pivotal role in studies of Majorana zero modes. Spin polarization of a zero-bias peak (ZBP) is used to determine its topological nature. Here we investigate the YSR states of single magnetic impurities at the surface of Pb using the fully relativistic first-principles simulations including band structure of Pb and five 3$d$ orbitals of the impurity in the superconducting state. We show that for single Fe and Co impurities, strong SOC can induce a ZBP with rotation of the impurity magnetic moment and that the ZBP has large spin polarization in contrast to effective model studies. Conditions for a ZBP from a single magnetic impurity are discussed. Our results are relevant to longer atomic chains considering their canting and noncollinear magnetism.


\end{abstract}


\maketitle






{\noindent {\it Introduction}}~~~
When magnetic impurities are present at surfaces of superconductors, quasiparticle excitations can appear within superconducting (SC) gaps due to exchange coupling between the impurity magnetic moments and the conduction electron spins. These excitations referred to as Yu-Shiba-Rusinov (YSR) states \cite{Yu1965,Shiba1968,Rusinov1969} have drawn lots of attention due to ongoing search for Majorana zero-energy modes (MZMs)
\cite{Kitaev2001,Alicea2012,Sarma2015} in long ferromagnetic atomic chains at $s$-wave superconductors with strong spin-orbit coupling (SOC) \cite{Klinovaja2013,Pientka2013,Vazifeh2013,Christensen2016,JLi2014,Ruby2015,JLi2018,NadjPerge2014,Jeon2017,Feldman2017,Pawlak2016,Ruby2017,Cornils2017,Schneider2021-2,Schneider2021-3,Arrachea2021}.
Magnetic properties of adatoms at superconductors were shown to be measured using YSR states in scanning tunneling microscopy/spectroscopy (STM/S) with much higher resolution \cite{Schneider2021}. YSR states can be also used to understand pairing symmetries \cite{Salkola1997,Balatsky2006,Shindou2010,Zitko2010}.

So far, YSR states have been mostly studied using effective models
\cite{Yu1965,Shiba1968,Rusinov1969,Morr2003,Flatte2000,Vazifeh2013,Christensen2016,Klinovaja2013,Pientka2013,JLi2014,Ruby2015,Ruby2015-2,JLi2018,Salkola1997,Balatsky2006,Kotetes2015}
based on single orbitals without realistic band structures and/or spin-orbit coupling (SOC) of SC substrates.
In long ferromagnetic atomic chains, zero-energy bound states, i.e. zero-bias peaks (ZBPs), arising from the ends of the chains are interpreted as topological MZMs when they have particle-hole symmetry and significant spin polarization, while ZBPs with zero spin polarization are considered to be topologically trivial \cite{Jeon2017,Feldman2017,JLi2018}. For long Fe atomic chains on SC Pb, a spin-polarized ZBP was experimentally observed only at the ends \cite{NadjPerge2014,Ruby2015,Pawlak2016,Jeon2017,Feldman2017}, whereas that was not the case for long Co chains \cite{Ruby2017}. Single interstitial Fe impurities at SC Fe(Te,Se) revealed zero-energy bound states \cite{JYin2015} without
an external magnetic field, and their origins are under active study \cite{KJiang2019,PFan2021,DFWang2021}. Given the intriguing results, first-principles studies of YSR states in the presence of SOC would shed light into the search for MZMs in a wider
range of systems.


As a first step toward achieving the goal, we study the YSR states of a single Fe, Co, or Mn magnetic impurity adsorbed on SC Pb, using first-principles simulations in the SC state. With strong SOC in Pb, we find that the YSR states are greatly affected by the rotation of the Fe or Co moment and that even a ZBP with large (normalized) spin polarization can occur at certain directions, although effective models \cite{Jeon2017,Feldman2017,JLi2018} suggest zero normalized spin polarization for trivial ZBPs. Our results clearly 
show new important features arising from strong SOC when realistic band structure and all 3$d$ orbitals are considered.


{\noindent{\it Methods}}~~~We simulate the normal state of Pb using fully the relativistic screened Korringa-Kohn-Rostoker (SKKR) Green's function method \cite{Csire2015} within density-functional theory (DFT). We employ the embedded cluster method \cite{Lazarovits2002} self-consistently within the SKKR formalism to a single Fe/Co/Mn magnetic impurity at the surface of the semi-infinite Pb(110) slab.
Assuming a $s$-wave effective pairing potential, we solve the Dirac-Bogoliubov-de Gennes (DBdG) equations for the self-consistently obtained 
normal-state heterostructure \cite{Csire2018,Saunderson2020}. We set the SC gap $\Delta$ of Pb (magnetic atom) to be the experimental value, 
1.36~meV \cite{Lykken1971} (zero). For method details, refer to Ref.~\cite{Nyari2021} and the Supplemental Material (SM).


\begin{figure*}[!tbhp]
\begin{center}
\includegraphics[width=0.95\textwidth]{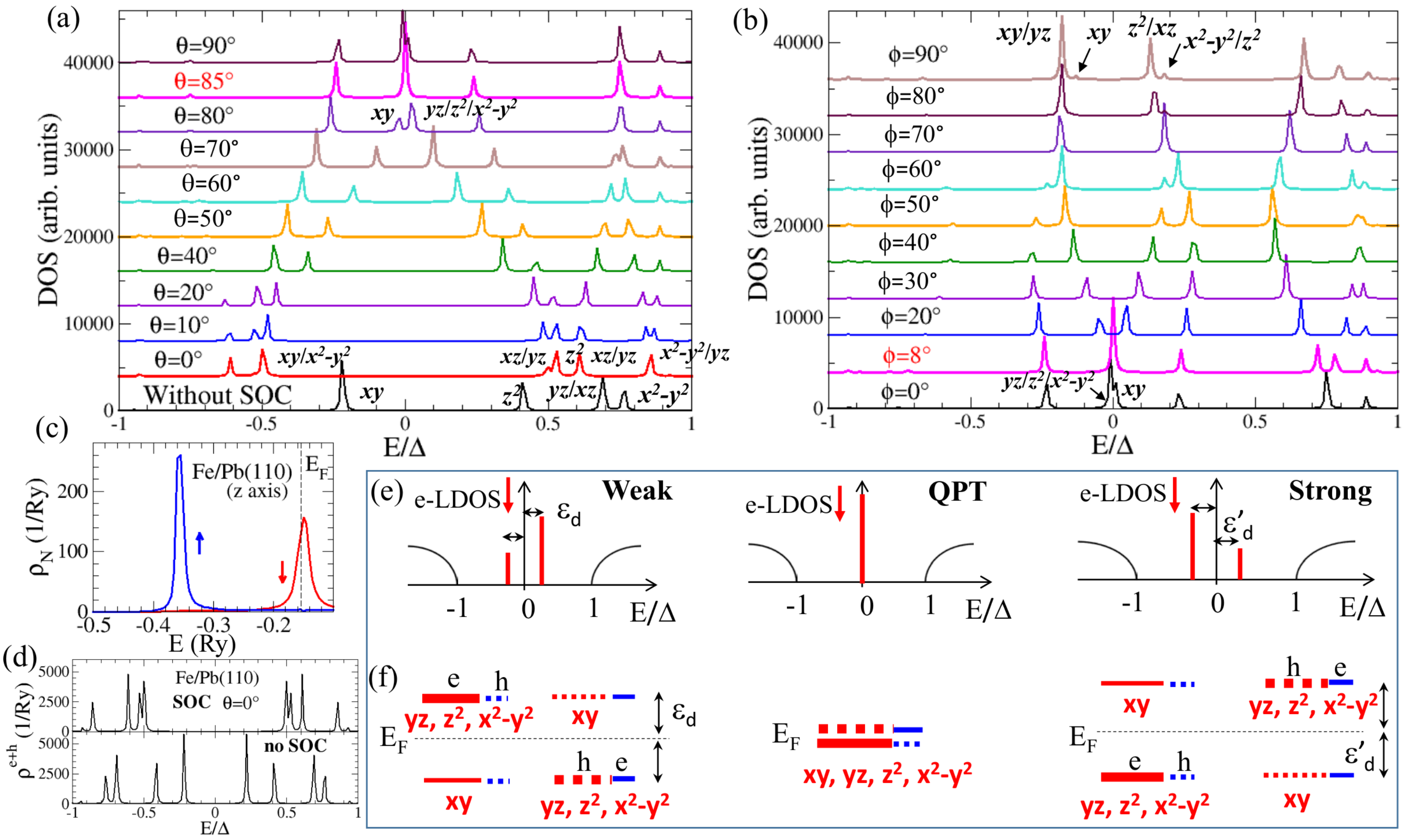}
\caption{(a) Fe {\it electron} part of the LDOS (sum of spin-up and spin-down) as a function of $\theta$ (angle of the Fe moment) in the $xz$ plane with SOC in the SC state. Only the bottom plot is obtained without SOC. (b) Fe electron LDOS (sum of spin-up and spin-down) as a function of $\phi$ in the $xy$ plane with SOC in the SC state. In (a) and (b), the plots (zero energy: $E_{\rm F}$) are vertically shifted for clarity and the QPT angles 
are shown as red. (c) Normal-state Fe spin-up and spin-down DOS with SOC (dashed line: $E_{\rm F}$). The $x$ ($y$) axis is along the 
[$\overline{1}$10] ([001]) direction. The $z$ axis is normal to the surface. The coordinates for the orbital decomposition shown in (a) and (b) do 
not change with the moment rotation. (d) Total Fe LDOS from electron and hole parts with and without SOC.
(e) Schematic diagram of merging of the deepest spin-down {\it electron} YSR pair at $E_{\rm F}$ and swapping as the moment rotates, and (f)
corresponding orbital characters in each case where electron (solid), hole (dashed), spin-up (blue), spin-down (red). Line thickness and length in
(f) are proportional to the LDOS magnitude. In (a), (b), and (f) corresponding $d$ orbitals are indicated.}
\label{fig:Fe-Pb}
\end{center}
\end{figure*}

{\noindent {\it Fe impurity on Pb(110)}}~~~In the normal state, for an Fe impurity on Pb, the Fe spin-up density of states (DOS) is fully occupied 
and the Fermi level $E_{\rm F}$ crosses slightly below the spin-down DOS peak [Fig.~\ref{fig:Fe-Pb}(c)]. Thus, there is large Fe spin-down DOS at $E_{\rm F}$, while the Fe spin-up states contribute very little at $E_{\rm F}$. The Fe DOS spectra are broadened due to
significant hybridization between the Fe impurity and the Pb substrate. The calculated spin and orbital moments are listed in Table~\ref{tab:1}.
The magnetic exchange energy is obtained from the difference between the Fe spin-up DOS peak and spin-down DOS peak energies (Table~\ref{tab:1}).

Let us first discuss the local density of states (LDOS) in the SC state without SOC. The calculated total Fe LDOS from electron and hole parts is symmetric with respect to $E_{\rm F}$ due to particle-hole symmetry, as shown in the bottom panel of Fig.~\ref{fig:Fe-Pb}(d), where the Fe moment points normal to the surface, i.e., along the $z$ axis. We confirm that without SOC, the LDOS does not depend on the Fe moment direction.
We henceforth present only {\it electron} part of the impurity LDOS unless specified otherwise. As shown in the bottom plot of Fig.~\ref{fig:Fe-Pb}(a), we find five pairs of {\it electron} Fe LDOS peaks at $E/\Delta=\pm0.22$, $\pm$0.41, $\pm$0.69, $\pm$0.77, and $\pm$0.94, where only one of each pair is dominant. The dominant peaks are all spin-down polarized except for the last pair, while the weak peaks
are spin-up polarized (Fig.~S1 in the SM). (The peak at $-$0.94$\Delta$ and the weak peaks are visible in Fig.S1.) The YSR peaks at $-$0.22$\Delta$
and 0.41$\Delta$ originate from $d_{xy}$ and $d_{z^2}$ orbitals, respectively, while the peaks at 0.69$\Delta$ and 0.77$\Delta$ arise from $d_{xz,yz}$ and $d_{x^2-y^2}$ orbitals, respectively. The peak at $-$0.94$\Delta$ originate from $s$, $d_{z^2}$, and $d_{x^2-y^2}$ orbitals.

\begin{table}[!h]
\centering
\caption{Calculated normal-state spin and orbital moments of the impurity, magnetic exchange energy, and existence of ZBP for the moment rotation. 
SOC is included. The number next to ZBP indicates the number of ZBPs with the rotation.}
\begin{tabular}{c|c|c|c} \hline \hline
                &  Fe/Pb(110)   & Co/Pb(110)    & Mn/Pb(110)    \\ \hline
Spin moment     & 3.518 $\mu_B$ & 2.183 $\mu_B$ & 4.812 $\mu_B$ \\
Orbital moment  & 0.906 $\mu_B$ & 1.215 $\mu_B$ & 0.027 $\mu_B$ \\
Exchange energy & 0.205 Ry      & 0.145 Ry      & 0.275 Ry      \\
$xz$-plane rotation   &    ZBP (1)    & ZBP (2)       & No ZBP        \\
$xy$-plane rotation   &    ZBP (1)    & No ZBP        & No ZBP        \\  \hline \hline
\end{tabular}
\label{tab:1}
\end{table}

Now we turn on SOC in the Fe impurity and the SC substrate, and we rotate the Fe moment with angle $\theta$ from the $z$ axis in the $xz$ plane and with angle $\phi$ from the $x$ axis in the $xy$ plane. The $x$ ($y$) axis is along the [$\overline{1}$10] ([001]) direction of Pb. Here the $x$, $y$, and $z$ axes indicate global coordinates where the $z$ axis is always normal to the surface despite the Fe moment rotation. Figure~\ref{fig:Fe-Pb}(a) and (b) show the {\it electron} part (sum of spin-up and spin-down) of the LDOS spectra as a function of $\theta$ and $\phi$, respectively
(see the top panel of Fig.~\ref{fig:Fe-Pb}(d) for the total LDOS with SOC.)
For orbital decomposition we use the global coordinates, while for spin polarization, the total or electron LDOS is decomposed into parallel (spin-up) and antiparallel (spin-down) components of the rotating Fe moment direction. Due to the two mirror symmetry planes, $1/4$ of the $xy$ plane and $xz$ plane suffices to study the effect of moment rotation. The LDOS spectra with SOC greatly differ from that without SOC.
With SOC, the energies and characters of YSR pairs change a lot as the Fe moment rotates. SOC allows mixing of 3$d$ orbitals with different magnetic quantum numbers, $m_l$, even for the Fe moment normal to the surface. With $\theta=0$, the
deepest YSR pair appears at $-0.5\Delta$ and $+0.5\Delta$ arising from \{$d_{xy}$,$d_{x^2-y^2}$\} and \{$d_{yz}$,$d_{xz}$,$d_{z^2}$\}, respectively.
When the moment points along the $x$ axis, the deepest YSR pair occurs at $-0.01\Delta$ and $+0.01\Delta$ originating from \{$d_{xy}$\} and \{$d_{yz}$,$d_{z^2}$,$d_{x^2-y^2}$\}, respectively. As the moment direction becomes closer to the $y$ axis, the deepest YSR pair is not well
isolated from the other YSR states and two deep YSR pairs start to hybridize.

\begin{figure}[!bht]
\begin{center}
\includegraphics[width=0.45\textwidth]{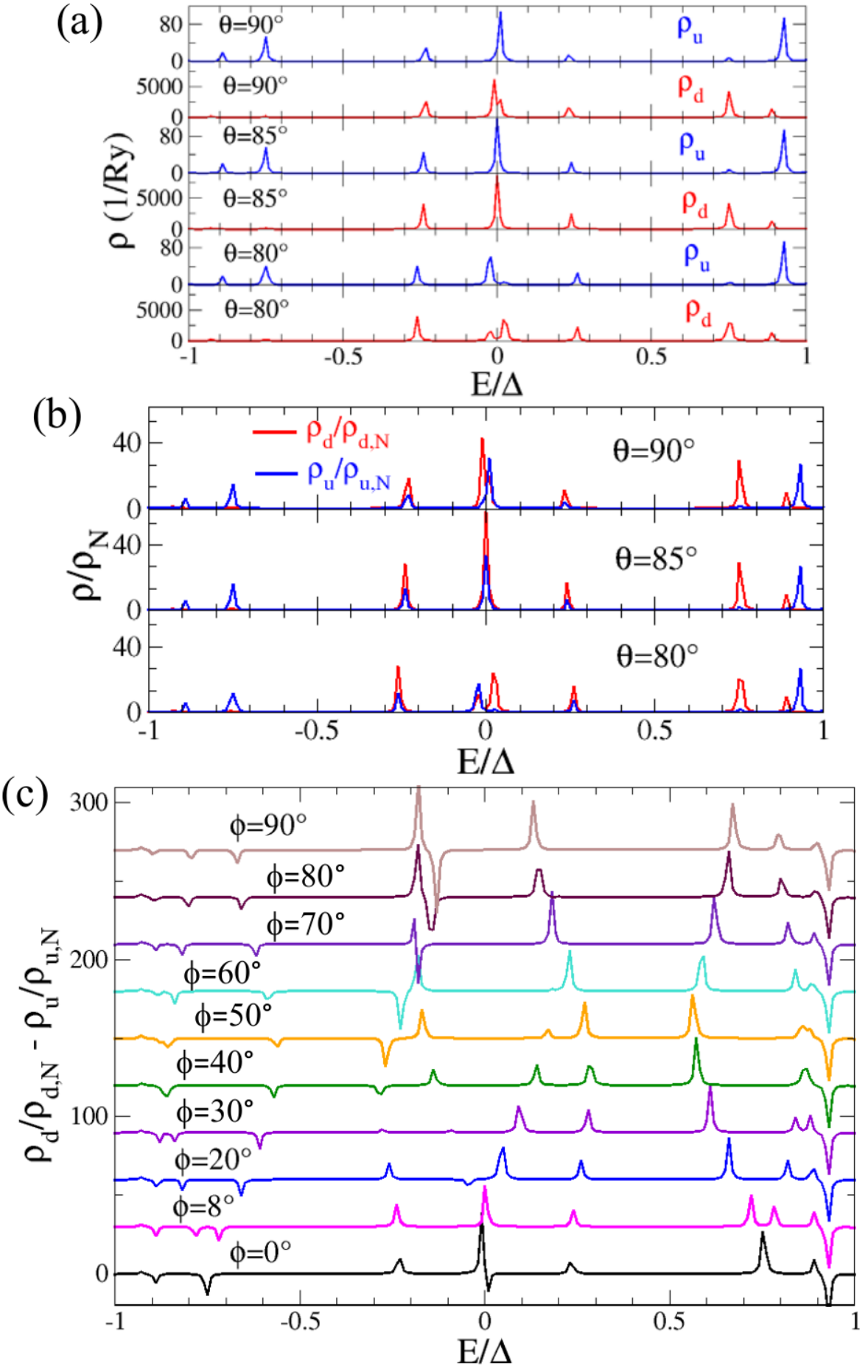}
\caption{(a) Fe {\it electron} spin-polarized LDOS in the SC state near the QPT. (b) The LDOS in (a) normalized with respect to the normal-state spin-polarized DOS. (c) Difference of the normalized Fe spin-down and spin-up electron LDOS as a function of $\phi$.}
\label{fig:Fe-spin}
\end{center}
\end{figure}

Interestingly, with SOC, in most cases, each electron part of the YSR pair within $\pm0.5\Delta$ has comparable spin-down DOS at both positive and negative energies (Figs.~\ref{fig:Fe-spin}(a) and~S2). Suppose that without SOC, the \{$e$,$\downarrow$;$h$,$\uparrow$\} component of a $d$ orbital contributes to the positive-energy YSR state. Strong SOC like Pb allows significant spin mixing, resulting in a significant contribution of the \{$e$,$\uparrow$;$h$,$\downarrow$\} component to the positive-energy YSR state. Particle-hole symmetry dictates that this component is equivalent to \{$e$,$\downarrow$;$h$,$\uparrow$\} at the negative-energy YSR state, such that a large
(small) spin-down LDOS peak at the positive (negative) energy illustrated in the leftmost scheme of Fig.~\ref{fig:Fe-Pb}(e) appears.

Importantly, we observe that a deepest YSR pair (with substantial spin-down DOS at positive and negative energies) crosses $E_{\rm F}$, or equivalently forms a ZBP, as a function of $\theta$ and $\phi$. For the $xz$-plane rotation, as $\theta$ increases, the positive-energy and negative-energy peaks
of the pair start to merge near $\theta=85^{\circ}$ and then they are swapped [Fig.~\ref{fig:Fe-Pb}(a)]. The crossing and swapping of the peaks are
schematically shown in Fig.~\ref{fig:Fe-Pb}(e) and (f). For the $xy$-plane rotation, a ZBP is formed near $\phi=8^{\circ}$ [Fig.~\ref{fig:Fe-Pb}(b)].
When the two peaks of the pair share the same $m_l$ values, they are likely to merge at $E_{\rm F}$. For example, at $\theta=80^{\circ}$, $d_{xy}$ orbital at the negative energy and $d_{x^2-y^2}$ orbital at the positive energy have $m_l=\pm2$ components.

This YSR pair crossing $E_{\rm F}$ is similar to a quantum phase transition (QPT) expected from SOC-free effective models
\cite{Sakurai1970,Salkola1997,Ruby2015-2,Farinacci2018,Heinrich2018}. In these models, when an exchange coupling between the impurity spin and conduction electron spin exceeds a critical value (i.e., strong coupling regime), the spin-unpolarized SC state becomes unstable such that the expectation value of the $z$ component of conduction electron spin at the impurity site changes to 1/2. The case that the exchange coupling is less than the critical value is referred to as weak coupling regime. At the critical exchange coupling, the YSR state appears at
$E_{\rm F}$, i.e., ZBP, with equal contributions from electron and hole parts. The QPT was observed for magnetic impurities or molecules \cite{Cornils2017,Hatter2015,Farinacci2018,HHuang2020,HDing2021}.
Our scenario qualitatively differs from them since the ZBP in our case is induced by the rotation of the impurity moment. The QPT or ZBP induced by the moment rotation in the presence of SOC has not been studied for realistic or experimental systems, although there is an $l=1$ effective-model study \cite{Kim2015}. The conditions for a ZBP from a single magnetic impurity are discussed later.

Apparently, the spin-down electron LDOS is always dominant over the spin-up electron LDOS at the ZBP as well as at both positive and negative energies of the deepest YSR pair except for near $\phi=90^{\circ}$ (Figs.~\ref{fig:Fe-spin}(a) and S2). This trend of spin polarization is also found for the
{\it hole} part of the LDOS (Fig.~S3). Note that spin projections can be experimentally measured even in the presence of SOC \cite{Jeon2017,Feldman2017,CHLi2014,ZXie2014}. We normalize the spin-polarized SC-state LDOS ($\rho_{\rm d}$,$\rho_{\rm u}$)
over normal-state spin-polarized LDOS ($\rho_{{\rm d},{\rm N}}$,$\rho_{{\rm u},{\rm N}}$), in order to provide relevance to
STM/S experiments and to compare with effective models \cite{JLi2018}. The conductance difference between spin-down and spin-up STM tip polarizations
is proportional to a difference between the normalized spin-down and spin-up electron LDOS,
$\Delta\rho$ ($=\rho_{\rm d}/\rho_{{\rm d},{\rm N}} - \rho_{\rm u}/\rho_{{\rm u},{\rm N}}$) \cite{JLi2018}. 
At $\theta=80^{\circ}$ the {\it normalized} spin-up (spin-down) deepest YSR state appears at the negative (positive) energy, while at $\theta=90^{\circ}$ the normalized spin polarization is reversed [Fig.~\ref{fig:Fe-spin}(b)]. With strong SOC like Pb, there is a substantial normalized spin-down contribution even when the normalized spin-up LDOS is dominant, in contrast to the cases of weak SOC and without SOC 
(Fig.S4). When the ZBP appears at $\theta=85^{\circ}$, we find large normalized spin polarization. Overall, $\Delta\rho$ shows complex dependencies on energy and rotation angles ($\theta$, $\phi$) [Figs.~\ref{fig:Fe-spin}(c), S5]. 


\begin{figure}[!hbt]
\begin{center}
\includegraphics[width=0.48\textwidth]{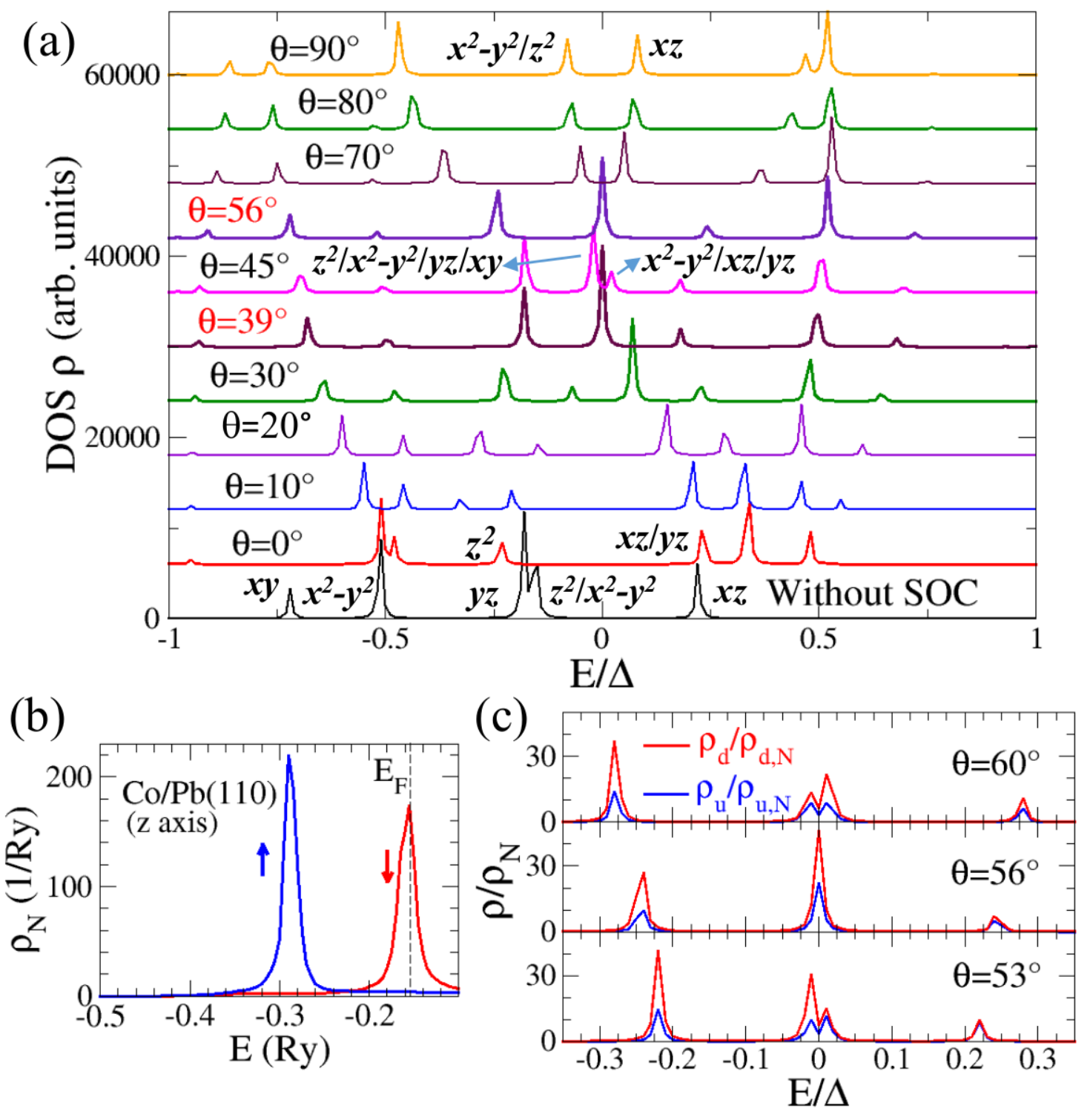}
\caption{(a) Co electron (sum of spin-up and spin-down) LDOS as a function of $\theta$ with SOC in the SC state. Only the bottom plot is obtained without SOC. (b) Normal-state Co spin-up and spin-down DOS with SOC (dashed line: $E_{\rm F}$). (c) Spin-polarized Co electron SC-state
LDOS normalized with respect to the normal-state spin-polarized DOS near the QPT.}
\label{fig:Co-Pb}
\end{center}
\end{figure}

{\noindent {\it Co impurity on Pb(110)}}~~~In the normal state, for a Co impurity on Pb, the Co spin-down DOS at $E_{\rm F}$ is higher than that for the Fe case [Fig.~\ref{fig:Co-Pb}(b)]. As shown at the bottom of Fig.~\ref{fig:Co-Pb}(a), without SOC, we find five pairs of LDOS peaks from $d_{xy}$, $d_{x^2-y^2}$, $d_{yz}$, \{$d_{z^2}$,$d_{x^2-y^2}$\}, and $d_{xz}$ orbitals, respectively. (Only five dominant peaks are visible due to the scale.) This shows that different types of magnetic impurities provide very different YSR states even for the same SC substrate. With SOC, the energies of the YSR states change a lot as a function of $\theta$ and $\phi$. Similarly to the Fe case, for
$-0.5\Delta < E < 0.5\Delta$, the YSR pairs show dominant spin-down peaks at positive and negative energies [Figs.~\ref{fig:Co-Pb}(c),~S6]. However,
the angular dependence qualitatively differs from that of the Fe case. The merging of the deepest YSR pair occurs twice, i.e., near $\theta=39^{\circ}$ and 56$^{\circ}$, in the $xz$-plane rotation [Fig.~\ref{fig:Co-Pb}(a)]. For the $xy$-plane rotation, however, we do not find a ZBP [Fig.~S7(b)]. The difference between the Co and Fe cases is attributed to the fact that the Co spin (orbital) moment is much smaller (larger) than the Fe spin (orbital) moment (Table~\ref{tab:1}). Our first-principles simulations show strong correlation among the normal-state LDOS, the impurity spin and orbital moments, and their effects on the YSR states. Despite the different LDOS features, the spin polarization of the ZBPs near $\theta=39^{\circ}$ and 56$^{\circ}$ is large whether it is normalized by the normal-state LDOS or not [Figs.~\ref{fig:Co-Pb}(c), S7(a)]. The split peaks of the deepest YSR 
pair are now spin-down at both negative and positive energies near the QPT angles.

\begin{figure}[!hbt]
\begin{center}
\includegraphics[width=0.48\textwidth]{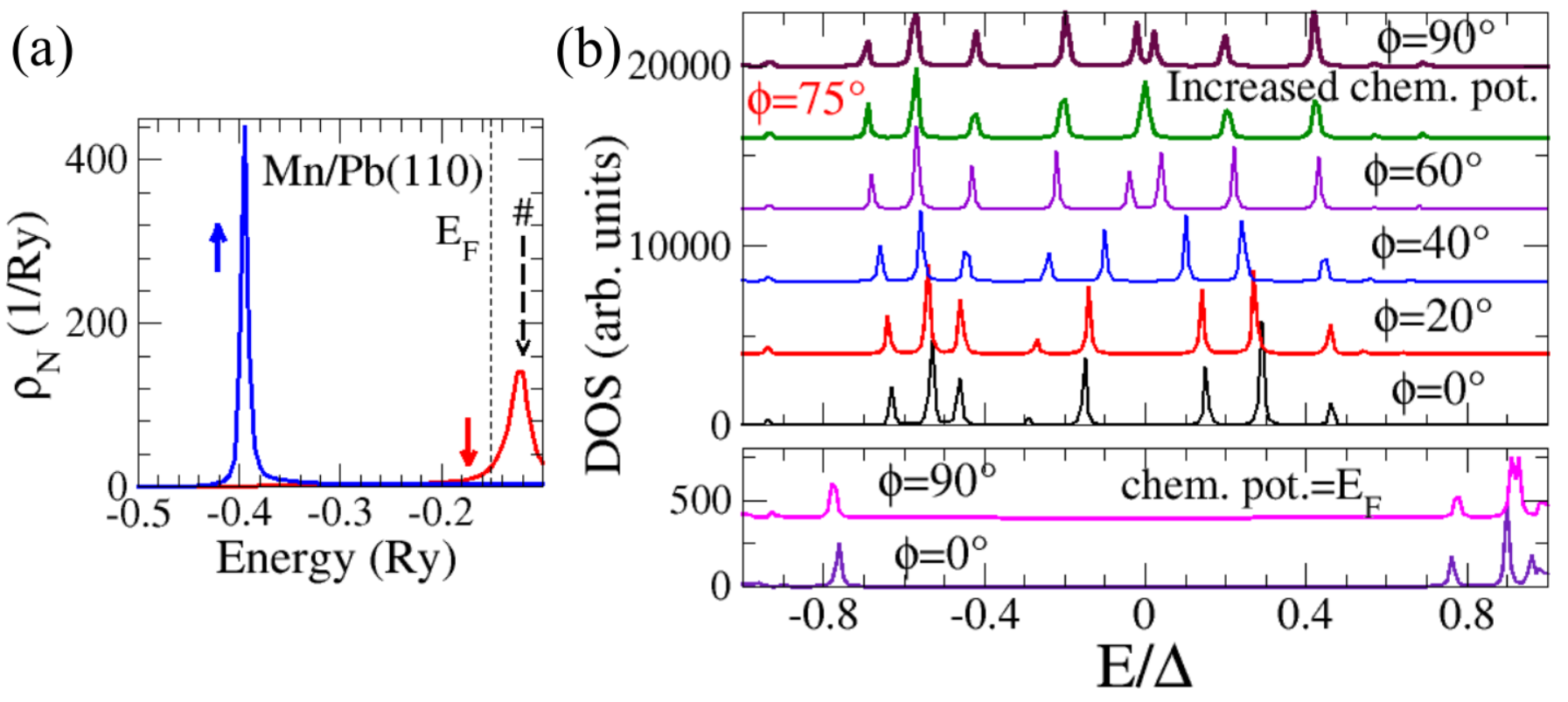}
\caption{(a) Normal-state Mn spin-up and spin-down DOS with SOC (dashed line: $E_{\rm F}$). (b) Electron Mn SC-state LDOS (sum of spin-up and spin-down) as a function of $\phi$ with SOC when the chemical potential is $E_{\rm F}$ (bottom panel) and is at the thick dashed arrow denoted in (a) (top panel).}
\label{fig:Mn-Pb}
\end{center}
\end{figure}

{\noindent {\it Mn impurity on Pb(110)}}~~~In the normal state, for a Mn impurity on Pb, the Mn spin-down DOS is barely occupied since $E_{\rm F}$ crosses the tail of the Mn spin-down DOS [Fig.~\ref{fig:Mn-Pb}(a)]. In the SC state, even with SOC, the electron LDOS peaks appear only near the gap edges and they do not change much with the moment rotation in the $xz$ and $xy$ planes [bottom panel in Fig.~\ref{fig:Mn-Pb}(b)]. This is due to the low normal-state Mn DOS at $E_{\rm F}$, although the Mn spin moment is the largest among Fe, Co, Mn impurities (Table~\ref{tab:1}). In order to show importance of large normal-state impurity DOS, we calculate the SC-state Mn LDOS as a function of $\phi$ when the chemical potential is raised to the spin-down LDOS peak marked in Fig.~\ref{fig:Mn-Pb}(a). We find that the YSR states now significantly change with $\phi$ and that even a ZBP appears near $\phi=75^{\circ}$.

{\noindent{\it Discussion}}~~~
From our first-principles simulations, we extract the following features common to the ZBP cases. First, the normal-state impurity DOS is high at $E_{\rm F}$. Second, it is more likely to observe a ZBP for a SC substrate with strong SOC like Pb.
Third, the impurity magnetic moment has substantial in-plane components which may be related to Rashba SOC of the SC substrate.
Overall, we find strong interplay between the impurity DOS at $E_{\rm F}$ and the magnetic moment of the impurity, resulting in the intriguing,
complex dependence of the YSR states on the impurity moment direction in the presence of SOC.

Since spin polarization is one of the key ingredients to determine the topological nature of ZBPs, we compare our results with effective model
studies \cite{JLi2018}, although we do not expect MZMs from single magnetic impurities. In the single-orbital effective models \cite{JLi2018},
$\Delta\rho$ is an antisymmetric function of energy relative to $E_{\rm F}$ due to a sum rule applied for simplicity. Thus, when a YSR pair merge to form a ZBP, the normalized spin polarization of the ZBP vanishes. Now when the band structure of the SC substrate and all five 3$d$ orbitals of the magnetic impurities are taken into account, $\Delta\rho$ is {\it not} an antisymmetric function of energy with or without SOC.
In the presence of strong SOC, both the height and sign of $\Delta\rho$ for the
deep YSR pair strongly depend on the moment direction. Thus, the ZBP in our case has large (normalized) spin polarization.
\\

{\noindent {\it Conclusion}}~~~
We simulated single magnetic impurities at the surface of SC Pb(110) by solving the DBdG equations using the embedded cluster method within the SKKR formalism. DFT band structures and multiple 3$d$ orbitals of the impurities were included in the normal and SC states. For the Fe and Co impurities, we found a ZBP with large (normalized) spin polarization fir the rotation of the impurity moment, whereas for the Mn impurity, a ZBP was not found due to the low impurity DOS at $E_{\rm F}$. Our results suggest importance of including realistic band structures and multiple orbitals in studies of YSR states of magnetic impurities and they are also relevant to understand effects of canting and noncollinear magnetism on the YSR states including ZBPs in long magnetic atomic chains. Furthermore, they imply our need to search for signatures of topological MZMs beyond spin polarization.
\\


{\noindent {\it Acknowledgements}}~~~
The computational support was provided by the Virginia Tech Advanced Research Computing as well as the Extreme Science and Engineering Discovery Environment (XSEDE) under Project number DMR060009N (K.P.) funded by the United States National Science Foundation Grant number ACI-1548562.
B.U., L.S., B.N., and A.L. were supported by the Hungarian National Research, Development and Innovation Office under Contract number K138628
and through the NRDI Fund TKP2020 698 IES (Grant number BME-IE-NAT). The authors are grateful to discussion with Gabor Csire, Yi-Ting Hsu,
and Sangjun Jeon.

\clearpage


\begin{thebibliography}{50}%
\makeatletter
\providecommand \@ifxundefined [1]{%
 \@ifx{#1\undefined}
}%
\providecommand \@ifnum [1]{%
 \ifnum #1\expandafter \@firstoftwo
 \else \expandafter \@secondoftwo
 \fi
}%
\providecommand \@ifx [1]{%
 \ifx #1\expandafter \@firstoftwo
 \else \expandafter \@secondoftwo
 \fi
}%
\providecommand \natexlab [1]{#1}%
\providecommand \enquote  [1]{``#1''}%
\providecommand \bibnamefont  [1]{#1}%
\providecommand \bibfnamefont [1]{#1}%
\providecommand \citenamefont [1]{#1}%
\providecommand \href@noop [0]{\@secondoftwo}%
\providecommand \href [0]{\begingroup \@sanitize@url \@href}%
\providecommand \@href[1]{\@@startlink{#1}\@@href}%
\providecommand \@@href[1]{\endgroup#1\@@endlink}%
\providecommand \@sanitize@url [0]{\catcode `\\12\catcode `\$12\catcode
  `\&12\catcode `\#12\catcode `\^12\catcode `\_12\catcode `\%12\relax}%
\providecommand \@@startlink[1]{}%
\providecommand \@@endlink[0]{}%
\providecommand \url  [0]{\begingroup\@sanitize@url \@url }%
\providecommand \@url [1]{\endgroup\@href {#1}{\urlprefix }}%
\providecommand \urlprefix  [0]{URL }%
\providecommand \Eprint [0]{\href }%
\providecommand \doibase [0]{https://doi.org/}%
\providecommand \selectlanguage [0]{\@gobble}%
\providecommand \bibinfo  [0]{\@secondoftwo}%
\providecommand \bibfield  [0]{\@secondoftwo}%
\providecommand \translation [1]{[#1]}%
\providecommand \BibitemOpen [0]{}%
\providecommand \bibitemStop [0]{}%
\providecommand \bibitemNoStop [0]{.\EOS\space}%
\providecommand \EOS [0]{\spacefactor3000\relax}%
\providecommand \BibitemShut  [1]{\csname bibitem#1\endcsname}%
\let\auto@bib@innerbib\@empty
\bibitem [{\citenamefont {Yu}(1965)}]{Yu1965}%
  \BibitemOpen
  \bibfield  {author} {\bibinfo {author} {\bibfnamefont {L.}~\bibnamefont
  {Yu}},\ }\bibfield  {title} {\bibinfo {title} {{Bound State in
  Superconductors with Paramagnetic Impurities}},\ }\href
  {https://wulixb.iphy.ac.cn/en/article/id/851} {\bibfield  {journal} {\bibinfo
   {journal} {Acta Physica Sinica}\ }\textbf {\bibinfo {volume} {21}},\
  \bibinfo {pages} {75} (\bibinfo {year} {1965})}\BibitemShut {NoStop}%
\bibitem [{\citenamefont {Shiba}(1968)}]{Shiba1968}%
  \BibitemOpen
  \bibfield  {author} {\bibinfo {author} {\bibfnamefont {H.}~\bibnamefont
  {Shiba}},\ }\bibfield  {title} {\bibinfo {title} {{Classical Spins in
  Superconductors}},\ }\href {https://doi.org/10.1143/PTP.40.435} {\bibfield
  {journal} {\bibinfo  {journal} {Progress of Theoretical Physics}\ }\textbf
  {\bibinfo {volume} {40}},\ \bibinfo {pages} {435} (\bibinfo {year} {1968})},\
  \Eprint
  {https://arxiv.org/abs/https://academic.oup.com/ptp/article-pdf/40/3/435/5185550/40-3-435.pdf}
  {https://academic.oup.com/ptp/article-pdf/40/3/435/5185550/40-3-435.pdf}
  \BibitemShut {NoStop}%
\bibitem [{\citenamefont {Rusinov}(1969)}]{Rusinov1969}%
  \BibitemOpen
  \bibfield  {author} {\bibinfo {author} {\bibfnamefont {A.~I.}\ \bibnamefont
  {Rusinov}},\ }\bibfield  {title} {\bibinfo {title} {{Superconductivity near a
  paramagnetic impurity}},\ }\href {https://www.osti.gov/biblio/7348743}
  {\bibfield  {journal} {\bibinfo  {journal} {JETP Lett. (USSR) (Engl. Transl.)
  (United States)}\ }\textbf {\bibinfo {volume} {9}},\ \bibinfo {pages} {85}
  (\bibinfo {year} {1969})}\BibitemShut {NoStop}%
\bibitem [{\citenamefont {Kitaev}(2001)}]{Kitaev2001}%
  \BibitemOpen
  \bibfield  {author} {\bibinfo {author} {\bibfnamefont {A.~Y.}\ \bibnamefont
  {Kitaev}},\ }\bibfield  {title} {\bibinfo {title} {{Unpaired Majorana
  fermions in quantum wires}},\ }\href@noop {} {\bibfield  {journal} {\bibinfo
  {journal} {Phys.-Usp.}\ }\textbf {\bibinfo {volume} {44}},\ \bibinfo {pages}
  {131} (\bibinfo {year} {2001})}\BibitemShut {NoStop}%
\bibitem [{\citenamefont {Alicea}(2012)}]{Alicea2012}%
  \BibitemOpen
  \bibfield  {author} {\bibinfo {author} {\bibfnamefont {J.}~\bibnamefont
  {Alicea}},\ }\bibfield  {title} {\bibinfo {title} {{New directions in the
  pursuit of Majorana fermions in solid state systems}},\ }\href@noop {}
  {\bibfield  {journal} {\bibinfo  {journal} {Rep. Prog. Phys.}\ }\textbf
  {\bibinfo {volume} {75}},\ \bibinfo {pages} {076501} (\bibinfo {year}
  {2012})}\BibitemShut {NoStop}%
\bibitem [{\citenamefont {Sarma}\ \emph {et~al.}(2015)\citenamefont {Sarma},
  \citenamefont {Freedman},\ and\ \citenamefont {Nayak}}]{Sarma2015}%
  \BibitemOpen
  \bibfield  {author} {\bibinfo {author} {\bibfnamefont {S.~D.}\ \bibnamefont
  {Sarma}}, \bibinfo {author} {\bibfnamefont {M.}~\bibnamefont {Freedman}},\
  and\ \bibinfo {author} {\bibfnamefont {C.}~\bibnamefont {Nayak}},\ }\bibfield
   {title} {\bibinfo {title} {{Majorana zero modes and topological quantum
  computation}},\ }\href {https://doi.org/10.1038/npjqi.2015.1} {\bibfield
  {journal} {\bibinfo  {journal} {npj Quantum Information}\ }\textbf {\bibinfo
  {volume} {1}},\ \bibinfo {pages} {15001} (\bibinfo {year}
  {2015})}\BibitemShut {NoStop}%
\bibitem [{\citenamefont {Klinovaja}\ \emph {et~al.}(2013)\citenamefont
  {Klinovaja}, \citenamefont {Stano}, \citenamefont {Yazdani},\ and\
  \citenamefont {Loss}}]{Klinovaja2013}%
  \BibitemOpen
  \bibfield  {author} {\bibinfo {author} {\bibfnamefont {J.}~\bibnamefont
  {Klinovaja}}, \bibinfo {author} {\bibfnamefont {P.}~\bibnamefont {Stano}},
  \bibinfo {author} {\bibfnamefont {A.}~\bibnamefont {Yazdani}},\ and\ \bibinfo
  {author} {\bibfnamefont {D.}~\bibnamefont {Loss}},\ }\bibfield  {title}
  {\bibinfo {title} {{Topological Superconductivity and Majorana Fermions in
  RKKY Systems}},\ }\href {https://doi.org/10.1103/PhysRevLett.111.186805}
  {\bibfield  {journal} {\bibinfo  {journal} {Phys. Rev. Lett.}\ }\textbf
  {\bibinfo {volume} {111}},\ \bibinfo {pages} {186805} (\bibinfo {year}
  {2013})}\BibitemShut {NoStop}%
\bibitem [{\citenamefont {Pientka}\ \emph {et~al.}(2013)\citenamefont
  {Pientka}, \citenamefont {Glazman},\ and\ \citenamefont {von
  Oppen}}]{Pientka2013}%
  \BibitemOpen
  \bibfield  {author} {\bibinfo {author} {\bibfnamefont {F.}~\bibnamefont
  {Pientka}}, \bibinfo {author} {\bibfnamefont {L.~I.}\ \bibnamefont
  {Glazman}},\ and\ \bibinfo {author} {\bibfnamefont {F.}~\bibnamefont {von
  Oppen}},\ }\bibfield  {title} {\bibinfo {title} {{Topological superconducting
  phase in helical Shiba chains}},\ }\href
  {https://doi.org/10.1103/PhysRevB.88.155420} {\bibfield  {journal} {\bibinfo
  {journal} {Phys. Rev. B}\ }\textbf {\bibinfo {volume} {88}},\ \bibinfo
  {pages} {155420} (\bibinfo {year} {2013})}\BibitemShut {NoStop}%
\bibitem [{\citenamefont {Vazifeh}\ and\ \citenamefont
  {Franz}(2013)}]{Vazifeh2013}%
  \BibitemOpen
  \bibfield  {author} {\bibinfo {author} {\bibfnamefont {M.~M.}\ \bibnamefont
  {Vazifeh}}\ and\ \bibinfo {author} {\bibfnamefont {M.}~\bibnamefont
  {Franz}},\ }\bibfield  {title} {\bibinfo {title} {Self-organized topological
  state with majorana fermions},\ }\href
  {https://doi.org/10.1103/PhysRevLett.111.206802} {\bibfield  {journal}
  {\bibinfo  {journal} {Phys. Rev. Lett.}\ }\textbf {\bibinfo {volume} {111}},\
  \bibinfo {pages} {206802} (\bibinfo {year} {2013})}\BibitemShut {NoStop}%
\bibitem [{\citenamefont {Christensen}\ \emph {et~al.}(2016)\citenamefont
  {Christensen}, \citenamefont {Schecter}, \citenamefont {Flensberg},
  \citenamefont {Andersen},\ and\ \citenamefont {Paaske}}]{Christensen2016}%
  \BibitemOpen
  \bibfield  {author} {\bibinfo {author} {\bibfnamefont {M.~H.}\ \bibnamefont
  {Christensen}}, \bibinfo {author} {\bibfnamefont {M.}~\bibnamefont
  {Schecter}}, \bibinfo {author} {\bibfnamefont {K.}~\bibnamefont {Flensberg}},
  \bibinfo {author} {\bibfnamefont {B.~M.}\ \bibnamefont {Andersen}},\ and\
  \bibinfo {author} {\bibfnamefont {J.}~\bibnamefont {Paaske}},\ }\bibfield
  {title} {\bibinfo {title} {{Spiral magnetic order and topological
  superconductivity in a chain of magnetic adatoms on a two-dimensional
  superconductor}},\ }\href {https://doi.org/10.1103/PhysRevB.94.144509}
  {\bibfield  {journal} {\bibinfo  {journal} {Phys. Rev. B}\ }\textbf {\bibinfo
  {volume} {94}},\ \bibinfo {pages} {144509} (\bibinfo {year}
  {2016})}\BibitemShut {NoStop}%
\bibitem [{\citenamefont {Li}\ \emph {et~al.}(2014{\natexlab{a}})\citenamefont
  {Li}, \citenamefont {Chen}, \citenamefont {Drozdov}, \citenamefont {Yazdani},
  \citenamefont {Bernevig},\ and\ \citenamefont {MacDonald}}]{JLi2014}%
  \BibitemOpen
  \bibfield  {author} {\bibinfo {author} {\bibfnamefont {J.}~\bibnamefont
  {Li}}, \bibinfo {author} {\bibfnamefont {H.}~\bibnamefont {Chen}}, \bibinfo
  {author} {\bibfnamefont {I.~K.}\ \bibnamefont {Drozdov}}, \bibinfo {author}
  {\bibfnamefont {A.}~\bibnamefont {Yazdani}}, \bibinfo {author} {\bibfnamefont
  {B.~A.}\ \bibnamefont {Bernevig}},\ and\ \bibinfo {author} {\bibfnamefont
  {A.~H.}\ \bibnamefont {MacDonald}},\ }\bibfield  {title} {\bibinfo {title}
  {Topological superconductivity induced by ferromagnetic metal chains},\
  }\href {https://doi.org/10.1103/PhysRevB.90.235433} {\bibfield  {journal}
  {\bibinfo  {journal} {Phys. Rev. B}\ }\textbf {\bibinfo {volume} {90}},\
  \bibinfo {pages} {235433} (\bibinfo {year} {2014}{\natexlab{a}})}\BibitemShut
  {NoStop}%
\bibitem [{\citenamefont {Ruby}\ \emph
  {et~al.}(2015{\natexlab{a}})\citenamefont {Ruby}, \citenamefont {Pientka},
  \citenamefont {Peng}, \citenamefont {von Oppen}, \citenamefont {Heinrich},\
  and\ \citenamefont {Franke}}]{Ruby2015}%
  \BibitemOpen
  \bibfield  {author} {\bibinfo {author} {\bibfnamefont {M.}~\bibnamefont
  {Ruby}}, \bibinfo {author} {\bibfnamefont {F.}~\bibnamefont {Pientka}},
  \bibinfo {author} {\bibfnamefont {Y.}~\bibnamefont {Peng}}, \bibinfo {author}
  {\bibfnamefont {F.}~\bibnamefont {von Oppen}}, \bibinfo {author}
  {\bibfnamefont {B.~W.}\ \bibnamefont {Heinrich}},\ and\ \bibinfo {author}
  {\bibfnamefont {K.~J.}\ \bibnamefont {Franke}},\ }\bibfield  {title}
  {\bibinfo {title} {{End States and Subgap Structure in Proximity-Coupled
  Chains of Magnetic Adatoms}},\ }\href
  {https://doi.org/10.1103/PhysRevLett.115.197204} {\bibfield  {journal}
  {\bibinfo  {journal} {Phys. Rev. Lett.}\ }\textbf {\bibinfo {volume} {115}},\
  \bibinfo {pages} {197204} (\bibinfo {year} {2015}{\natexlab{a}})}\BibitemShut
  {NoStop}%
\bibitem [{\citenamefont {Li}\ \emph {et~al.}(2018)\citenamefont {Li},
  \citenamefont {Jeon}, \citenamefont {Xie}, \citenamefont {Yazdani},\ and\
  \citenamefont {Bernevig}}]{JLi2018}%
  \BibitemOpen
  \bibfield  {author} {\bibinfo {author} {\bibfnamefont {J.}~\bibnamefont
  {Li}}, \bibinfo {author} {\bibfnamefont {S.}~\bibnamefont {Jeon}}, \bibinfo
  {author} {\bibfnamefont {Y.}~\bibnamefont {Xie}}, \bibinfo {author}
  {\bibfnamefont {A.}~\bibnamefont {Yazdani}},\ and\ \bibinfo {author}
  {\bibfnamefont {B.~A.}\ \bibnamefont {Bernevig}},\ }\bibfield  {title}
  {\bibinfo {title} {{Majorana spin in magnetic atomic chain systems}},\ }\href
  {https://doi.org/10.1103/PhysRevB.97.125119} {\bibfield  {journal} {\bibinfo
  {journal} {Phys. Rev. B}\ }\textbf {\bibinfo {volume} {97}},\ \bibinfo
  {pages} {125119} (\bibinfo {year} {2018})}\BibitemShut {NoStop}%
\bibitem [{\citenamefont {Nadj-Perge}\ \emph {et~al.}(2014)\citenamefont
  {Nadj-Perge}, \citenamefont {Drozdov}, \citenamefont {Li}, \citenamefont
  {Chen}, \citenamefont {Jeon}, \citenamefont {Seo}, \citenamefont {MacDonald},
  \citenamefont {Bernevig},\ and\ \citenamefont {Yazdani}}]{NadjPerge2014}%
  \BibitemOpen
  \bibfield  {author} {\bibinfo {author} {\bibfnamefont {S.}~\bibnamefont
  {Nadj-Perge}}, \bibinfo {author} {\bibfnamefont {I.~K.}\ \bibnamefont
  {Drozdov}}, \bibinfo {author} {\bibfnamefont {J.}~\bibnamefont {Li}},
  \bibinfo {author} {\bibfnamefont {H.}~\bibnamefont {Chen}}, \bibinfo {author}
  {\bibfnamefont {S.}~\bibnamefont {Jeon}}, \bibinfo {author} {\bibfnamefont
  {J.}~\bibnamefont {Seo}}, \bibinfo {author} {\bibfnamefont {A.~H.}\
  \bibnamefont {MacDonald}}, \bibinfo {author} {\bibfnamefont {B.~A.}\
  \bibnamefont {Bernevig}},\ and\ \bibinfo {author} {\bibfnamefont
  {A.}~\bibnamefont {Yazdani}},\ }\bibfield  {title} {\bibinfo {title}
  {{Observation of Majorana fermions in ferromagnetic atomic chains on a
  superconductor}},\ }\href@noop {} {\bibfield  {journal} {\bibinfo  {journal}
  {Science}\ }\textbf {\bibinfo {volume} {346}},\ \bibinfo {pages} {602}
  (\bibinfo {year} {2014})}\BibitemShut {NoStop}%
\bibitem [{\citenamefont {Jeon}\ \emph {et~al.}(2017)\citenamefont {Jeon},
  \citenamefont {Xie}, \citenamefont {Li}, \citenamefont {Wang}, \citenamefont
  {Bernevig},\ and\ \citenamefont {Yazdani}}]{Jeon2017}%
  \BibitemOpen
  \bibfield  {author} {\bibinfo {author} {\bibfnamefont {S.}~\bibnamefont
  {Jeon}}, \bibinfo {author} {\bibfnamefont {Y.}~\bibnamefont {Xie}}, \bibinfo
  {author} {\bibfnamefont {J.}~\bibnamefont {Li}}, \bibinfo {author}
  {\bibfnamefont {Z.}~\bibnamefont {Wang}}, \bibinfo {author} {\bibfnamefont
  {B.~A.}\ \bibnamefont {Bernevig}},\ and\ \bibinfo {author} {\bibfnamefont
  {A.}~\bibnamefont {Yazdani}},\ }\bibfield  {title} {\bibinfo {title}
  {{Distinguishing a Majorana zero mode using spin-resolved measurements}},\
  }\href {https://doi.org/10.1126/science.aan3670} {\bibfield  {journal}
  {\bibinfo  {journal} {Science}\ }\textbf {\bibinfo {volume} {358}},\ \bibinfo
  {pages} {772} (\bibinfo {year} {2017})}\BibitemShut {NoStop}%
\bibitem [{\citenamefont {Feldman}\ \emph {et~al.}(2017)\citenamefont
  {Feldman}, \citenamefont {Randeria}, \citenamefont {Li}, \citenamefont
  {Jeon}, \citenamefont {Xie}, \citenamefont {Wang}, \citenamefont {Drozdov},
  \citenamefont {Andrei~Bernevig},\ and\ \citenamefont
  {Yazdani}}]{Feldman2017}%
  \BibitemOpen
  \bibfield  {author} {\bibinfo {author} {\bibfnamefont {B.~E.}\ \bibnamefont
  {Feldman}}, \bibinfo {author} {\bibfnamefont {M.~T.}\ \bibnamefont
  {Randeria}}, \bibinfo {author} {\bibfnamefont {J.}~\bibnamefont {Li}},
  \bibinfo {author} {\bibfnamefont {S.}~\bibnamefont {Jeon}}, \bibinfo {author}
  {\bibfnamefont {Y.}~\bibnamefont {Xie}}, \bibinfo {author} {\bibfnamefont
  {Z.}~\bibnamefont {Wang}}, \bibinfo {author} {\bibfnamefont {I.~K.}\
  \bibnamefont {Drozdov}}, \bibinfo {author} {\bibfnamefont {B.}~\bibnamefont
  {Andrei~Bernevig}},\ and\ \bibinfo {author} {\bibfnamefont {A.}~\bibnamefont
  {Yazdani}},\ }\bibfield  {title} {\bibinfo {title} {{High-resolution studies
  of the Majorana atomic chain platform}},\ }\href@noop {} {\bibfield
  {journal} {\bibinfo  {journal} {Nat. Phys.}\ }\textbf {\bibinfo {volume}
  {13}},\ \bibinfo {pages} {286} (\bibinfo {year} {2017})}\BibitemShut
  {NoStop}%
\bibitem [{\citenamefont {Pawlak}\ \emph {et~al.}(2016)\citenamefont {Pawlak},
  \citenamefont {Kisiel}, \citenamefont {Klinovaja}, \citenamefont {Meier},
  \citenamefont {Kawai}, \citenamefont {Glatzel}, \citenamefont {Loss},\ and\
  \citenamefont {Meyer}}]{Pawlak2016}%
  \BibitemOpen
  \bibfield  {author} {\bibinfo {author} {\bibfnamefont {R.}~\bibnamefont
  {Pawlak}}, \bibinfo {author} {\bibfnamefont {M.}~\bibnamefont {Kisiel}},
  \bibinfo {author} {\bibfnamefont {J.}~\bibnamefont {Klinovaja}}, \bibinfo
  {author} {\bibfnamefont {T.}~\bibnamefont {Meier}}, \bibinfo {author}
  {\bibfnamefont {S.}~\bibnamefont {Kawai}}, \bibinfo {author} {\bibfnamefont
  {T.}~\bibnamefont {Glatzel}}, \bibinfo {author} {\bibfnamefont
  {D.}~\bibnamefont {Loss}},\ and\ \bibinfo {author} {\bibfnamefont
  {E.}~\bibnamefont {Meyer}},\ }\bibfield  {title} {\bibinfo {title} {{Probing
  atomic structure and Majorana wavefunctions in mono-atomic Fe chains on
  superconducting Pb surface}},\ }\href {https://doi.org/10.1038/npjqi.2016.35}
  {\bibfield  {journal} {\bibinfo  {journal} {npj Quantum Information}\
  }\textbf {\bibinfo {volume} {2}},\ \bibinfo {pages} {16035} (\bibinfo {year}
  {2016})}\BibitemShut {NoStop}%
\bibitem [{\citenamefont {Ruby}\ \emph {et~al.}(2017)\citenamefont {Ruby},
  \citenamefont {Heinrich}, \citenamefont {Peng}, \citenamefont {von Oppen},\
  and\ \citenamefont {Franke}}]{Ruby2017}%
  \BibitemOpen
  \bibfield  {author} {\bibinfo {author} {\bibfnamefont {M.}~\bibnamefont
  {Ruby}}, \bibinfo {author} {\bibfnamefont {B.~W.}\ \bibnamefont {Heinrich}},
  \bibinfo {author} {\bibfnamefont {Y.}~\bibnamefont {Peng}}, \bibinfo {author}
  {\bibfnamefont {F.}~\bibnamefont {von Oppen}},\ and\ \bibinfo {author}
  {\bibfnamefont {K.~J.}\ \bibnamefont {Franke}},\ }\bibfield  {title}
  {\bibinfo {title} {{Exploring a Proximity-Coupled Co Chain on Pb(110) as a
  Possible Majorana Platform}},\ }\href@noop {} {\bibfield  {journal} {\bibinfo
   {journal} {Nano Lett.}\ }\textbf {\bibinfo {volume} {17}},\ \bibinfo {pages}
  {4473} (\bibinfo {year} {2017})}\BibitemShut {NoStop}%
\bibitem [{\citenamefont {Cornils}\ \emph {et~al.}(2017)\citenamefont
  {Cornils}, \citenamefont {Kamlapure}, \citenamefont {Zhou}, \citenamefont
  {Pradhan}, \citenamefont {Khajetoorians}, \citenamefont {Fransson},
  \citenamefont {Wiebe},\ and\ \citenamefont {Wiesendanger}}]{Cornils2017}%
  \BibitemOpen
  \bibfield  {author} {\bibinfo {author} {\bibfnamefont {L.}~\bibnamefont
  {Cornils}}, \bibinfo {author} {\bibfnamefont {A.}~\bibnamefont {Kamlapure}},
  \bibinfo {author} {\bibfnamefont {L.}~\bibnamefont {Zhou}}, \bibinfo {author}
  {\bibfnamefont {S.}~\bibnamefont {Pradhan}}, \bibinfo {author} {\bibfnamefont
  {A.~A.}\ \bibnamefont {Khajetoorians}}, \bibinfo {author} {\bibfnamefont
  {J.}~\bibnamefont {Fransson}}, \bibinfo {author} {\bibfnamefont
  {J.}~\bibnamefont {Wiebe}},\ and\ \bibinfo {author} {\bibfnamefont
  {R.}~\bibnamefont {Wiesendanger}},\ }\bibfield  {title} {\bibinfo {title}
  {Spin-resolved spectroscopy of the yu-shiba-rusinov states of individual
  atoms},\ }\href {https://doi.org/10.1103/PhysRevLett.119.197002} {\bibfield
  {journal} {\bibinfo  {journal} {Phys. Rev. Lett.}\ }\textbf {\bibinfo
  {volume} {119}},\ \bibinfo {pages} {197002} (\bibinfo {year}
  {2017})}\BibitemShut {NoStop}%
\bibitem [{\citenamefont {Schneider}\ \emph
  {et~al.}(2021{\natexlab{a}})\citenamefont {Schneider}, \citenamefont {Beck},
  \citenamefont {Neuhaus-Steinmetz}, \citenamefont {Posske}, \citenamefont
  {Wiebe},\ and\ \citenamefont {Wiesendanger}}]{Schneider2021-2}%
  \BibitemOpen
  \bibfield  {author} {\bibinfo {author} {\bibfnamefont {L.}~\bibnamefont
  {Schneider}}, \bibinfo {author} {\bibfnamefont {P.}~\bibnamefont {Beck}},
  \bibinfo {author} {\bibfnamefont {J.}~\bibnamefont {Neuhaus-Steinmetz}},
  \bibinfo {author} {\bibfnamefont {T.}~\bibnamefont {Posske}}, \bibinfo
  {author} {\bibfnamefont {J.}~\bibnamefont {Wiebe}},\ and\ \bibinfo {author}
  {\bibfnamefont {R.}~\bibnamefont {Wiesendanger}},\ }\href@noop {} {\bibinfo
  {title} {{Controlled length-dependent interaction of Majorana modes in
  Yu-Shiba-Rusinov chains}}} (\bibinfo {year} {2021}{\natexlab{a}}),\ \Eprint
  {https://arxiv.org/abs/2104.11503} {arXiv:2104.11503 [cond-mat.supr-con]}
  \BibitemShut {NoStop}%
\bibitem [{\citenamefont {Schneider}\ \emph
  {et~al.}(2021{\natexlab{b}})\citenamefont {Schneider}, \citenamefont {Beck},
  \citenamefont {Posske}, \citenamefont {Crawford}, \citenamefont {Mascot},
  \citenamefont {Rachel}, \citenamefont {Wiesendanger},\ and\ \citenamefont
  {Wiebe}}]{Schneider2021-3}%
  \BibitemOpen
  \bibfield  {author} {\bibinfo {author} {\bibfnamefont {L.}~\bibnamefont
  {Schneider}}, \bibinfo {author} {\bibfnamefont {P.}~\bibnamefont {Beck}},
  \bibinfo {author} {\bibfnamefont {T.}~\bibnamefont {Posske}}, \bibinfo
  {author} {\bibfnamefont {D.}~\bibnamefont {Crawford}}, \bibinfo {author}
  {\bibfnamefont {E.}~\bibnamefont {Mascot}}, \bibinfo {author} {\bibfnamefont
  {S.}~\bibnamefont {Rachel}}, \bibinfo {author} {\bibfnamefont
  {R.}~\bibnamefont {Wiesendanger}},\ and\ \bibinfo {author} {\bibfnamefont
  {J.}~\bibnamefont {Wiebe}},\ }\bibfield  {title} {\bibinfo {title}
  {{Topological Shiba bands in artificial spin chains on superconductors}},\
  }\bibfield  {journal} {\bibinfo  {journal} {Nature Physics}\ }\href
  {https://doi.org/10.1038/s41567-021-01234-y} {10.1038/s41567-021-01234-y}
  (\bibinfo {year} {2021}{\natexlab{b}})\BibitemShut {NoStop}%
\bibitem [{\citenamefont {Arrachea}(2021)}]{Arrachea2021}%
  \BibitemOpen
  \bibfield  {author} {\bibinfo {author} {\bibfnamefont {L.}~\bibnamefont
  {Arrachea}},\ }\href@noop {} {\bibinfo {title} {{Yu-Shiba-Rusinov multiplets
  and clusters of multiorbital adatoms in superconducting substrates: Subgap
  Green's function approach}}} (\bibinfo {year} {2021}),\ \Eprint
  {https://arxiv.org/abs/2108.02546} {arXiv:2108.02546 [cond-mat.mes-hall]}
  \BibitemShut {NoStop}%
\bibitem [{\citenamefont {Schneider}\ \emph
  {et~al.}(2021{\natexlab{c}})\citenamefont {Schneider}, \citenamefont {Beck},
  \citenamefont {Wiebe},\ and\ \citenamefont {Wiesendanger}}]{Schneider2021}%
  \BibitemOpen
  \bibfield  {author} {\bibinfo {author} {\bibfnamefont {L.}~\bibnamefont
  {Schneider}}, \bibinfo {author} {\bibfnamefont {P.}~\bibnamefont {Beck}},
  \bibinfo {author} {\bibfnamefont {J.}~\bibnamefont {Wiebe}},\ and\ \bibinfo
  {author} {\bibfnamefont {R.}~\bibnamefont {Wiesendanger}},\ }\bibfield
  {title} {\bibinfo {title} {Atomic-scale spin-polarization maps using
  functionalized superconducting probes},\ }\bibfield  {journal} {\bibinfo
  {journal} {Science Advances}\ }\textbf {\bibinfo {volume} {7}},\ \href
  {https://doi.org/10.1126/sciadv.abd7302} {10.1126/sciadv.abd7302} (\bibinfo
  {year} {2021}{\natexlab{c}})\BibitemShut {NoStop}%
\bibitem [{\citenamefont {Salkola}\ \emph {et~al.}(1997)\citenamefont
  {Salkola}, \citenamefont {Balatsky},\ and\ \citenamefont
  {Schrieffer}}]{Salkola1997}%
  \BibitemOpen
  \bibfield  {author} {\bibinfo {author} {\bibfnamefont {M.~I.}\ \bibnamefont
  {Salkola}}, \bibinfo {author} {\bibfnamefont {A.~V.}\ \bibnamefont
  {Balatsky}},\ and\ \bibinfo {author} {\bibfnamefont {J.~R.}\ \bibnamefont
  {Schrieffer}},\ }\bibfield  {title} {\bibinfo {title} {Spectral properties of
  quasiparticle excitations induced by magnetic moments in superconductors},\
  }\href {https://doi.org/10.1103/PhysRevB.55.12648} {\bibfield  {journal}
  {\bibinfo  {journal} {Phys. Rev. B}\ }\textbf {\bibinfo {volume} {55}},\
  \bibinfo {pages} {12648} (\bibinfo {year} {1997})}\BibitemShut {NoStop}%
\bibitem [{\citenamefont {Balatsky}\ \emph {et~al.}(2006)\citenamefont
  {Balatsky}, \citenamefont {Vekhter},\ and\ \citenamefont
  {Zhu}}]{Balatsky2006}%
  \BibitemOpen
  \bibfield  {author} {\bibinfo {author} {\bibfnamefont {A.~V.}\ \bibnamefont
  {Balatsky}}, \bibinfo {author} {\bibfnamefont {I.}~\bibnamefont {Vekhter}},\
  and\ \bibinfo {author} {\bibfnamefont {J.-X.}\ \bibnamefont {Zhu}},\
  }\bibfield  {title} {\bibinfo {title} {{Impurity-induced states in
  conventional and unconventional superconductors}},\ }\href
  {https://doi.org/10.1103/RevModPhys.78.373} {\bibfield  {journal} {\bibinfo
  {journal} {Rev. Mod. Phys.}\ }\textbf {\bibinfo {volume} {78}},\ \bibinfo
  {pages} {373} (\bibinfo {year} {2006})}\BibitemShut {NoStop}%
\bibitem [{\citenamefont {Shindou}\ \emph {et~al.}(2010)\citenamefont
  {Shindou}, \citenamefont {Furusaki},\ and\ \citenamefont
  {Nagaosa}}]{Shindou2010}%
  \BibitemOpen
  \bibfield  {author} {\bibinfo {author} {\bibfnamefont {R.}~\bibnamefont
  {Shindou}}, \bibinfo {author} {\bibfnamefont {A.}~\bibnamefont {Furusaki}},\
  and\ \bibinfo {author} {\bibfnamefont {N.}~\bibnamefont {Nagaosa}},\
  }\bibfield  {title} {\bibinfo {title} {{Quantum impurity spin in Majorana
  edge fermions}},\ }\href {https://doi.org/10.1103/PhysRevB.82.180505}
  {\bibfield  {journal} {\bibinfo  {journal} {Phys. Rev. B}\ }\textbf {\bibinfo
  {volume} {82}},\ \bibinfo {pages} {180505} (\bibinfo {year}
  {2010})}\BibitemShut {NoStop}%
\bibitem [{\citenamefont {\ifmmode~\check{Z}\else \v{Z}\fi{}itko}\ and\
  \citenamefont {Simon}(2011)}]{Zitko2010}%
  \BibitemOpen
  \bibfield  {author} {\bibinfo {author} {\bibfnamefont {R.}~\bibnamefont
  {\ifmmode~\check{Z}\else \v{Z}\fi{}itko}}\ and\ \bibinfo {author}
  {\bibfnamefont {P.}~\bibnamefont {Simon}},\ }\bibfield  {title} {\bibinfo
  {title} {{Quantum impurity coupled to Majorana edge fermions}},\ }\href
  {https://doi.org/10.1103/PhysRevB.84.195310} {\bibfield  {journal} {\bibinfo
  {journal} {Phys. Rev. B}\ }\textbf {\bibinfo {volume} {84}},\ \bibinfo
  {pages} {195310} (\bibinfo {year} {2011})}\BibitemShut {NoStop}%
\bibitem [{\citenamefont {Morr}\ and\ \citenamefont
  {Stavropoulos}(2003)}]{Morr2003}%
  \BibitemOpen
  \bibfield  {author} {\bibinfo {author} {\bibfnamefont {D.~K.}\ \bibnamefont
  {Morr}}\ and\ \bibinfo {author} {\bibfnamefont {N.~A.}\ \bibnamefont
  {Stavropoulos}},\ }\bibfield  {title} {\bibinfo {title} {{Quantum
  interference between impurities: Creating novel many-body states in s-wave
  superconductors}},\ }\href {https://doi.org/10.1103/PhysRevB.67.020502}
  {\bibfield  {journal} {\bibinfo  {journal} {Phys. Rev. B}\ }\textbf {\bibinfo
  {volume} {67}},\ \bibinfo {pages} {020502} (\bibinfo {year}
  {2003})}\BibitemShut {NoStop}%
\bibitem [{\citenamefont {Flatt\'e}\ and\ \citenamefont
  {Reynolds}(2000)}]{Flatte2000}%
  \BibitemOpen
  \bibfield  {author} {\bibinfo {author} {\bibfnamefont {M.~E.}\ \bibnamefont
  {Flatt\'e}}\ and\ \bibinfo {author} {\bibfnamefont {D.~E.}\ \bibnamefont
  {Reynolds}},\ }\bibfield  {title} {\bibinfo {title} {Local spectrum of a
  superconductor as a probe of interactions between magnetic impurities},\
  }\href {https://doi.org/10.1103/PhysRevB.61.14810} {\bibfield  {journal}
  {\bibinfo  {journal} {Phys. Rev. B}\ }\textbf {\bibinfo {volume} {61}},\
  \bibinfo {pages} {14810} (\bibinfo {year} {2000})}\BibitemShut {NoStop}%
\bibitem [{\citenamefont {Ruby}\ \emph
  {et~al.}(2015{\natexlab{b}})\citenamefont {Ruby}, \citenamefont {Pientka},
  \citenamefont {Peng}, \citenamefont {von Oppen}, \citenamefont {Heinrich},\
  and\ \citenamefont {Franke}}]{Ruby2015-2}%
  \BibitemOpen
  \bibfield  {author} {\bibinfo {author} {\bibfnamefont {M.}~\bibnamefont
  {Ruby}}, \bibinfo {author} {\bibfnamefont {F.}~\bibnamefont {Pientka}},
  \bibinfo {author} {\bibfnamefont {Y.}~\bibnamefont {Peng}}, \bibinfo {author}
  {\bibfnamefont {F.}~\bibnamefont {von Oppen}}, \bibinfo {author}
  {\bibfnamefont {B.~W.}\ \bibnamefont {Heinrich}},\ and\ \bibinfo {author}
  {\bibfnamefont {K.~J.}\ \bibnamefont {Franke}},\ }\bibfield  {title}
  {\bibinfo {title} {{Tunneling Processes into Localized Subgap States in
  Superconductors}},\ }\href {https://doi.org/10.1103/PhysRevLett.115.087001}
  {\bibfield  {journal} {\bibinfo  {journal} {Phys. Rev. Lett.}\ }\textbf
  {\bibinfo {volume} {115}},\ \bibinfo {pages} {087001} (\bibinfo {year}
  {2015}{\natexlab{b}})}\BibitemShut {NoStop}%
\bibitem [{\citenamefont {Kotetes}\ \emph {et~al.}(2015)\citenamefont
  {Kotetes}, \citenamefont {Mendler}, \citenamefont {Heimes},\ and\
  \citenamefont {Schön}}]{Kotetes2015}%
  \BibitemOpen
  \bibfield  {author} {\bibinfo {author} {\bibfnamefont {P.}~\bibnamefont
  {Kotetes}}, \bibinfo {author} {\bibfnamefont {D.}~\bibnamefont {Mendler}},
  \bibinfo {author} {\bibfnamefont {A.}~\bibnamefont {Heimes}},\ and\ \bibinfo
  {author} {\bibfnamefont {G.}~\bibnamefont {Schön}},\ }\bibfield  {title}
  {\bibinfo {title} {Majorana fermion fingerprints in spin-polarised scanning
  tunnelling microscopy},\ }\href@noop {} {\bibfield  {journal} {\bibinfo
  {journal} {Physica E: Low-dimensional Systems and Nanostructures}\ }\textbf
  {\bibinfo {volume} {74}},\ \bibinfo {pages} {614} (\bibinfo {year}
  {2015})}\BibitemShut {NoStop}%
\bibitem [{\citenamefont {Yin}\ \emph {et~al.}(2015)\citenamefont {Yin},
  \citenamefont {Wu}, \citenamefont {Wang}, \citenamefont {Ye}, \citenamefont
  {Gong}, \citenamefont {Hou}, \citenamefont {Shan}, \citenamefont {Li},
  \citenamefont {Liang}, \citenamefont {Wu}, \citenamefont {Li}, \citenamefont
  {Ting}, \citenamefont {Wang}, \citenamefont {Hu}, \citenamefont {Hor},
  \citenamefont {Ding},\ and\ \citenamefont {Pan}}]{JYin2015}%
  \BibitemOpen
  \bibfield  {author} {\bibinfo {author} {\bibfnamefont {J.-X.}\ \bibnamefont
  {Yin}}, \bibinfo {author} {\bibfnamefont {Z.}~\bibnamefont {Wu}}, \bibinfo
  {author} {\bibfnamefont {J.-H.}\ \bibnamefont {Wang}}, \bibinfo {author}
  {\bibfnamefont {Z.-Y.}\ \bibnamefont {Ye}}, \bibinfo {author} {\bibfnamefont
  {J.}~\bibnamefont {Gong}}, \bibinfo {author} {\bibfnamefont {X.-Y.}\
  \bibnamefont {Hou}}, \bibinfo {author} {\bibfnamefont {L.}~\bibnamefont
  {Shan}}, \bibinfo {author} {\bibfnamefont {A.}~\bibnamefont {Li}}, \bibinfo
  {author} {\bibfnamefont {X.-J.}\ \bibnamefont {Liang}}, \bibinfo {author}
  {\bibfnamefont {X.-X.}\ \bibnamefont {Wu}}, \bibinfo {author} {\bibfnamefont
  {J.}~\bibnamefont {Li}}, \bibinfo {author} {\bibfnamefont {C.-S.}\
  \bibnamefont {Ting}}, \bibinfo {author} {\bibfnamefont {Z.-Q.}\ \bibnamefont
  {Wang}}, \bibinfo {author} {\bibfnamefont {J.-P.}\ \bibnamefont {Hu}},
  \bibinfo {author} {\bibfnamefont {P.-H.}\ \bibnamefont {Hor}}, \bibinfo
  {author} {\bibfnamefont {H.}~\bibnamefont {Ding}},\ and\ \bibinfo {author}
  {\bibfnamefont {S.~H.}\ \bibnamefont {Pan}},\ }\bibfield  {title} {\bibinfo
  {title} {{Observation of a robust zero-energy bound state in iron-based
  superconductor Fe(Te,Se)}},\ }\href {https://doi.org/10.1038/nphys3371}
  {\bibfield  {journal} {\bibinfo  {journal} {Nature Physics}\ }\textbf
  {\bibinfo {volume} {11}},\ \bibinfo {pages} {543} (\bibinfo {year}
  {2015})}\BibitemShut {NoStop}%
\bibitem [{\citenamefont {Jiang}\ \emph {et~al.}(2019)\citenamefont {Jiang},
  \citenamefont {Dai},\ and\ \citenamefont {Wang}}]{KJiang2019}%
  \BibitemOpen
  \bibfield  {author} {\bibinfo {author} {\bibfnamefont {K.}~\bibnamefont
  {Jiang}}, \bibinfo {author} {\bibfnamefont {X.}~\bibnamefont {Dai}},\ and\
  \bibinfo {author} {\bibfnamefont {Z.}~\bibnamefont {Wang}},\ }\bibfield
  {title} {\bibinfo {title} {{Quantum Anomalous Vortex and Majorana Zero Mode
  in Iron-Based Superconductor Fe(Te,Se)}},\ }\href
  {https://doi.org/10.1103/PhysRevX.9.011033} {\bibfield  {journal} {\bibinfo
  {journal} {Phys. Rev. X}\ }\textbf {\bibinfo {volume} {9}},\ \bibinfo {pages}
  {011033} (\bibinfo {year} {2019})}\BibitemShut {NoStop}%
\bibitem [{\citenamefont {Fan}\ \emph {et~al.}(2021)\citenamefont {Fan},
  \citenamefont {Yang}, \citenamefont {Qian}, \citenamefont {Chen},
  \citenamefont {Zhang}, \citenamefont {Li}, \citenamefont {Huang},
  \citenamefont {Xing}, \citenamefont {Kong}, \citenamefont {Liu},
  \citenamefont {Jiang}, \citenamefont {Shen}, \citenamefont {Du},
  \citenamefont {Schneeloch}, \citenamefont {Zhong}, \citenamefont {Gu},
  \citenamefont {Wang}, \citenamefont {Ding},\ and\ \citenamefont
  {Gao}}]{PFan2021}%
  \BibitemOpen
  \bibfield  {author} {\bibinfo {author} {\bibfnamefont {P.}~\bibnamefont
  {Fan}}, \bibinfo {author} {\bibfnamefont {F.}~\bibnamefont {Yang}}, \bibinfo
  {author} {\bibfnamefont {G.}~\bibnamefont {Qian}}, \bibinfo {author}
  {\bibfnamefont {H.}~\bibnamefont {Chen}}, \bibinfo {author} {\bibfnamefont
  {Y.-Y.}\ \bibnamefont {Zhang}}, \bibinfo {author} {\bibfnamefont
  {G.}~\bibnamefont {Li}}, \bibinfo {author} {\bibfnamefont {Z.}~\bibnamefont
  {Huang}}, \bibinfo {author} {\bibfnamefont {Y.}~\bibnamefont {Xing}},
  \bibinfo {author} {\bibfnamefont {L.}~\bibnamefont {Kong}}, \bibinfo {author}
  {\bibfnamefont {W.}~\bibnamefont {Liu}}, \bibinfo {author} {\bibfnamefont
  {K.}~\bibnamefont {Jiang}}, \bibinfo {author} {\bibfnamefont
  {C.}~\bibnamefont {Shen}}, \bibinfo {author} {\bibfnamefont {S.}~\bibnamefont
  {Du}}, \bibinfo {author} {\bibfnamefont {J.}~\bibnamefont {Schneeloch}},
  \bibinfo {author} {\bibfnamefont {R.}~\bibnamefont {Zhong}}, \bibinfo
  {author} {\bibfnamefont {G.}~\bibnamefont {Gu}}, \bibinfo {author}
  {\bibfnamefont {Z.}~\bibnamefont {Wang}}, \bibinfo {author} {\bibfnamefont
  {H.}~\bibnamefont {Ding}},\ and\ \bibinfo {author} {\bibfnamefont {H.-J.}\
  \bibnamefont {Gao}},\ }\bibfield  {title} {\bibinfo {title} {{Observation of
  magnetic adatom-induced Majorana vortex and its hybridization with
  field-induced Majorana vortex in an iron-based superconductor}},\ }\href@noop
  {} {\bibfield  {journal} {\bibinfo  {journal} {Nature Communications}\
  }\textbf {\bibinfo {volume} {12}},\ \bibinfo {pages} {1348} (\bibinfo {year}
  {2021})}\BibitemShut {NoStop}%
\bibitem [{\citenamefont {Wang}\ \emph {et~al.}(2021)\citenamefont {Wang},
  \citenamefont {Wiebe}, \citenamefont {Zhong}, \citenamefont {Gu},\ and\
  \citenamefont {Wiesendanger}}]{DFWang2021}%
  \BibitemOpen
  \bibfield  {author} {\bibinfo {author} {\bibfnamefont {D.}~\bibnamefont
  {Wang}}, \bibinfo {author} {\bibfnamefont {J.}~\bibnamefont {Wiebe}},
  \bibinfo {author} {\bibfnamefont {R.}~\bibnamefont {Zhong}}, \bibinfo
  {author} {\bibfnamefont {G.}~\bibnamefont {Gu}},\ and\ \bibinfo {author}
  {\bibfnamefont {R.}~\bibnamefont {Wiesendanger}},\ }\bibfield  {title}
  {\bibinfo {title} {{Spin-Polarized Yu-Shiba-Rusinov States in an Iron-Based
  Superconductor}},\ }\href {https://doi.org/10.1103/PhysRevLett.126.076802}
  {\bibfield  {journal} {\bibinfo  {journal} {Phys. Rev. Lett.}\ }\textbf
  {\bibinfo {volume} {126}},\ \bibinfo {pages} {076802} (\bibinfo {year}
  {2021})}\BibitemShut {NoStop}%
\bibitem [{\citenamefont {Csire}\ \emph {et~al.}(2015)\citenamefont {Csire},
  \citenamefont {\'Ujfalussy}, \citenamefont {Cserti},\ and\ \citenamefont
  {Gy\ifmmode~\mbox{\H{o}}\else \H{o}\fi{}rffy}}]{Csire2015}%
  \BibitemOpen
  \bibfield  {author} {\bibinfo {author} {\bibfnamefont {G.}~\bibnamefont
  {Csire}}, \bibinfo {author} {\bibfnamefont {B.}~\bibnamefont {\'Ujfalussy}},
  \bibinfo {author} {\bibfnamefont {J.}~\bibnamefont {Cserti}},\ and\ \bibinfo
  {author} {\bibfnamefont {B.}~\bibnamefont {Gy\ifmmode~\mbox{\H{o}}\else
  \H{o}\fi{}rffy}},\ }\bibfield  {title} {\bibinfo {title} {{Multiple
  scattering theory for superconducting heterostructures}},\ }\href
  {https://doi.org/10.1103/PhysRevB.91.165142} {\bibfield  {journal} {\bibinfo
  {journal} {Phys. Rev. B}\ }\textbf {\bibinfo {volume} {91}},\ \bibinfo
  {pages} {165142} (\bibinfo {year} {2015})}\BibitemShut {NoStop}%
\bibitem [{\citenamefont {Lazarovits}\ \emph {et~al.}(2002)\citenamefont
  {Lazarovits}, \citenamefont {Szunyogh},\ and\ \citenamefont
  {Weinberger}}]{Lazarovits2002}%
  \BibitemOpen
  \bibfield  {author} {\bibinfo {author} {\bibfnamefont {B.}~\bibnamefont
  {Lazarovits}}, \bibinfo {author} {\bibfnamefont {L.}~\bibnamefont
  {Szunyogh}},\ and\ \bibinfo {author} {\bibfnamefont {P.}~\bibnamefont
  {Weinberger}},\ }\bibfield  {title} {\bibinfo {title} {{Fully relativistic
  calculation of magnetic properties of Fe, Co, and Ni adclusters on
  Ag(100)}},\ }\href {https://doi.org/10.1103/PhysRevB.65.104441} {\bibfield
  {journal} {\bibinfo  {journal} {Phys. Rev. B}\ }\textbf {\bibinfo {volume}
  {65}},\ \bibinfo {pages} {104441} (\bibinfo {year} {2002})}\BibitemShut
  {NoStop}%
\bibitem [{\citenamefont {Csire}\ \emph {et~al.}(2018)\citenamefont {Csire},
  \citenamefont {De\'ak}, \citenamefont {Ny\'ari}, \citenamefont {Ebert},
  \citenamefont {Annett},\ and\ \citenamefont {\'Ujfalussy}}]{Csire2018}%
  \BibitemOpen
  \bibfield  {author} {\bibinfo {author} {\bibfnamefont {G.}~\bibnamefont
  {Csire}}, \bibinfo {author} {\bibfnamefont {A.}~\bibnamefont {De\'ak}},
  \bibinfo {author} {\bibfnamefont {B.}~\bibnamefont {Ny\'ari}}, \bibinfo
  {author} {\bibfnamefont {H.}~\bibnamefont {Ebert}}, \bibinfo {author}
  {\bibfnamefont {J.~F.}\ \bibnamefont {Annett}},\ and\ \bibinfo {author}
  {\bibfnamefont {B.}~\bibnamefont {\'Ujfalussy}},\ }\bibfield  {title}
  {\bibinfo {title} {{Relativistic spin-polarized KKR theory for
  superconducting heterostructures: Oscillating order parameter in the Au layer
  of Nb/Au/Fe trilayers}},\ }\href {https://doi.org/10.1103/PhysRevB.97.024514}
  {\bibfield  {journal} {\bibinfo  {journal} {Phys. Rev. B}\ }\textbf {\bibinfo
  {volume} {97}},\ \bibinfo {pages} {024514} (\bibinfo {year}
  {2018})}\BibitemShut {NoStop}%
\bibitem [{\citenamefont {Saunderson}\ \emph {et~al.}(2020)\citenamefont
  {Saunderson}, \citenamefont {Gyorgyp\'al}, \citenamefont {Annett},
  \citenamefont {Csire}, \citenamefont {\'Ujfalussy},\ and\ \citenamefont
  {Gradhand}}]{Saunderson2020}%
  \BibitemOpen
  \bibfield  {author} {\bibinfo {author} {\bibfnamefont {T.~G.}\ \bibnamefont
  {Saunderson}}, \bibinfo {author} {\bibfnamefont {Z.}~\bibnamefont
  {Gyorgyp\'al}}, \bibinfo {author} {\bibfnamefont {J.~F.}\ \bibnamefont
  {Annett}}, \bibinfo {author} {\bibfnamefont {G.}~\bibnamefont {Csire}},
  \bibinfo {author} {\bibfnamefont {B.}~\bibnamefont {\'Ujfalussy}},\ and\
  \bibinfo {author} {\bibfnamefont {M.}~\bibnamefont {Gradhand}},\ }\bibfield
  {title} {\bibinfo {title} {{Real-space multiple scattering theory for
  superconductors with impurities}},\ }\href
  {https://doi.org/10.1103/PhysRevB.102.245106} {\bibfield  {journal} {\bibinfo
   {journal} {Phys. Rev. B}\ }\textbf {\bibinfo {volume} {102}},\ \bibinfo
  {pages} {245106} (\bibinfo {year} {2020})}\BibitemShut {NoStop}%
\bibitem [{\citenamefont {Lykken}\ \emph {et~al.}(1971)\citenamefont {Lykken},
  \citenamefont {Geiger}, \citenamefont {Dy},\ and\ \citenamefont
  {Mitchell}}]{Lykken1971}%
  \BibitemOpen
  \bibfield  {author} {\bibinfo {author} {\bibfnamefont {G.~I.}\ \bibnamefont
  {Lykken}}, \bibinfo {author} {\bibfnamefont {A.~L.}\ \bibnamefont {Geiger}},
  \bibinfo {author} {\bibfnamefont {K.~S.}\ \bibnamefont {Dy}},\ and\ \bibinfo
  {author} {\bibfnamefont {E.~N.}\ \bibnamefont {Mitchell}},\ }\bibfield
  {title} {\bibinfo {title} {{Measurement of the Superconducting Energy Gap and
  Fermi Velocity in Single-Crystal Lead Films by Electron Tunneling}},\ }\href
  {https://doi.org/10.1103/PhysRevB.4.1523} {\bibfield  {journal} {\bibinfo
  {journal} {Phys. Rev. B}\ }\textbf {\bibinfo {volume} {4}},\ \bibinfo {pages}
  {1523} (\bibinfo {year} {1971})}\BibitemShut {NoStop}%
\bibitem [{\citenamefont {Nyári}\ \emph {et~al.}(2021)\citenamefont {Nyári},
  \citenamefont {Lászlóffy}, \citenamefont {Szunyogh}, \citenamefont {Csire},
  \citenamefont {Park},\ and\ \citenamefont {Ujfalussy}}]{Nyari2021}%
  \BibitemOpen
  \bibfield  {author} {\bibinfo {author} {\bibfnamefont {B.}~\bibnamefont
  {Nyári}}, \bibinfo {author} {\bibfnamefont {A.}~\bibnamefont {Lászlóffy}},
  \bibinfo {author} {\bibfnamefont {L.}~\bibnamefont {Szunyogh}}, \bibinfo
  {author} {\bibfnamefont {G.}~\bibnamefont {Csire}}, \bibinfo {author}
  {\bibfnamefont {K.}~\bibnamefont {Park}},\ and\ \bibinfo {author}
  {\bibfnamefont {B.}~\bibnamefont {Ujfalussy}},\ }\href@noop {} {\bibinfo
  {title} {Relativistic first principles theory of yu--shiba--rusinov states
  applied to an mn adatom and mn dimers on nb(110)}} (\bibinfo {year} {2021}),\
  \Eprint {https://arxiv.org/abs/2109.03499} {arXiv:2109.03499
  [cond-mat.supr-con]} \BibitemShut {NoStop}%
\bibitem [{\citenamefont {Sakurai}(1970)}]{Sakurai1970}%
  \BibitemOpen
  \bibfield  {author} {\bibinfo {author} {\bibfnamefont {A.}~\bibnamefont
  {Sakurai}},\ }\bibfield  {title} {\bibinfo {title} {{Comments on
  Superconductors with Magnetic Impurities}},\ }\href
  {https://doi.org/10.1143/PTP.44.1472} {\bibfield  {journal} {\bibinfo
  {journal} {Progress of Theoretical Physics}\ }\textbf {\bibinfo {volume}
  {44}},\ \bibinfo {pages} {1472} (\bibinfo {year} {1970})}\BibitemShut
  {NoStop}%
\bibitem [{\citenamefont {Farinacci}\ \emph {et~al.}(2018)\citenamefont
  {Farinacci}, \citenamefont {Ahmadi}, \citenamefont {Reecht}, \citenamefont
  {Ruby}, \citenamefont {Bogdanoff}, \citenamefont {Peters}, \citenamefont
  {Heinrich}, \citenamefont {von Oppen},\ and\ \citenamefont
  {Franke}}]{Farinacci2018}%
  \BibitemOpen
  \bibfield  {author} {\bibinfo {author} {\bibfnamefont {L.}~\bibnamefont
  {Farinacci}}, \bibinfo {author} {\bibfnamefont {G.}~\bibnamefont {Ahmadi}},
  \bibinfo {author} {\bibfnamefont {G.}~\bibnamefont {Reecht}}, \bibinfo
  {author} {\bibfnamefont {M.}~\bibnamefont {Ruby}}, \bibinfo {author}
  {\bibfnamefont {N.}~\bibnamefont {Bogdanoff}}, \bibinfo {author}
  {\bibfnamefont {O.}~\bibnamefont {Peters}}, \bibinfo {author} {\bibfnamefont
  {B.~W.}\ \bibnamefont {Heinrich}}, \bibinfo {author} {\bibfnamefont
  {F.}~\bibnamefont {von Oppen}},\ and\ \bibinfo {author} {\bibfnamefont
  {K.~J.}\ \bibnamefont {Franke}},\ }\bibfield  {title} {\bibinfo {title}
  {{Tuning the Coupling of an Individual Magnetic Impurity to a Superconductor:
  Quantum Phase Transition and Transport}},\ }\href
  {https://doi.org/10.1103/PhysRevLett.121.196803} {\bibfield  {journal}
  {\bibinfo  {journal} {Phys. Rev. Lett.}\ }\textbf {\bibinfo {volume} {121}},\
  \bibinfo {pages} {196803} (\bibinfo {year} {2018})}\BibitemShut {NoStop}%
\bibitem [{\citenamefont {Heinrich}\ \emph {et~al.}(2018)\citenamefont
  {Heinrich}, \citenamefont {Pascual},\ and\ \citenamefont
  {Franke}}]{Heinrich2018}%
  \BibitemOpen
  \bibfield  {author} {\bibinfo {author} {\bibfnamefont {B.~W.}\ \bibnamefont
  {Heinrich}}, \bibinfo {author} {\bibfnamefont {J.~I.}\ \bibnamefont
  {Pascual}},\ and\ \bibinfo {author} {\bibfnamefont {K.~J.}\ \bibnamefont
  {Franke}},\ }\bibfield  {title} {\bibinfo {title} {Single magnetic adsorbates
  on s-wave superconductors},\ }\href@noop {} {\bibfield  {journal} {\bibinfo
  {journal} {Progress in Surface Science}\ }\textbf {\bibinfo {volume} {93}},\
  \bibinfo {pages} {1} (\bibinfo {year} {2018})}\BibitemShut {NoStop}%
\bibitem [{\citenamefont {Hatter}\ \emph {et~al.}(2015)\citenamefont {Hatter},
  \citenamefont {Heinrich}, \citenamefont {Ruby}, \citenamefont {Pascual},\
  and\ \citenamefont {Franke}}]{Hatter2015}%
  \BibitemOpen
  \bibfield  {author} {\bibinfo {author} {\bibfnamefont {N.}~\bibnamefont
  {Hatter}}, \bibinfo {author} {\bibfnamefont {B.~W.}\ \bibnamefont
  {Heinrich}}, \bibinfo {author} {\bibfnamefont {M.}~\bibnamefont {Ruby}},
  \bibinfo {author} {\bibfnamefont {J.~I.}\ \bibnamefont {Pascual}},\ and\
  \bibinfo {author} {\bibfnamefont {K.~J.}\ \bibnamefont {Franke}},\ }\bibfield
   {title} {\bibinfo {title} {{Magnetic anisotropy in Shiba bound states across
  a quantum phase transition}},\ }\href@noop {} {\bibfield  {journal} {\bibinfo
   {journal} {Nature Communications}\ }\textbf {\bibinfo {volume} {6}},\
  \bibinfo {pages} {8988} (\bibinfo {year} {2015})}\BibitemShut {NoStop}%
\bibitem [{\citenamefont {Huang}\ \emph {et~al.}(2020)\citenamefont {Huang},
  \citenamefont {Drost}, \citenamefont {Senkpiel}, \citenamefont {Padurariu},
  \citenamefont {Kubala}, \citenamefont {Yeyati}, \citenamefont {Cuevas},
  \citenamefont {Ankerhold}, \citenamefont {Kern},\ and\ \citenamefont
  {Ast}}]{HHuang2020}%
  \BibitemOpen
  \bibfield  {author} {\bibinfo {author} {\bibfnamefont {H.}~\bibnamefont
  {Huang}}, \bibinfo {author} {\bibfnamefont {R.}~\bibnamefont {Drost}},
  \bibinfo {author} {\bibfnamefont {J.}~\bibnamefont {Senkpiel}}, \bibinfo
  {author} {\bibfnamefont {C.}~\bibnamefont {Padurariu}}, \bibinfo {author}
  {\bibfnamefont {B.}~\bibnamefont {Kubala}}, \bibinfo {author} {\bibfnamefont
  {A.~L.}\ \bibnamefont {Yeyati}}, \bibinfo {author} {\bibfnamefont {J.~C.}\
  \bibnamefont {Cuevas}}, \bibinfo {author} {\bibfnamefont {J.}~\bibnamefont
  {Ankerhold}}, \bibinfo {author} {\bibfnamefont {K.}~\bibnamefont {Kern}},\
  and\ \bibinfo {author} {\bibfnamefont {C.~R.}\ \bibnamefont {Ast}},\
  }\bibfield  {title} {\bibinfo {title} {{Quantum phase transitions and the
  role of impurity-substrate hybridization in Yu-Shiba-Rusinov states}},\
  }\href@noop {} {\bibfield  {journal} {\bibinfo  {journal} {Communications
  Physics}\ }\textbf {\bibinfo {volume} {3}},\ \bibinfo {pages} {199} (\bibinfo
  {year} {2020})}\BibitemShut {NoStop}%
\bibitem [{\citenamefont {Ding}\ \emph {et~al.}(2021)\citenamefont {Ding},
  \citenamefont {Hu}, \citenamefont {Randeria}, \citenamefont {Hoffman},
  \citenamefont {Deb}, \citenamefont {Klinovaja}, \citenamefont {Loss},\ and\
  \citenamefont {Yazdani}}]{HDing2021}%
  \BibitemOpen
  \bibfield  {author} {\bibinfo {author} {\bibfnamefont {H.}~\bibnamefont
  {Ding}}, \bibinfo {author} {\bibfnamefont {Y.}~\bibnamefont {Hu}}, \bibinfo
  {author} {\bibfnamefont {M.~T.}\ \bibnamefont {Randeria}}, \bibinfo {author}
  {\bibfnamefont {S.}~\bibnamefont {Hoffman}}, \bibinfo {author} {\bibfnamefont
  {O.}~\bibnamefont {Deb}}, \bibinfo {author} {\bibfnamefont {J.}~\bibnamefont
  {Klinovaja}}, \bibinfo {author} {\bibfnamefont {D.}~\bibnamefont {Loss}},\
  and\ \bibinfo {author} {\bibfnamefont {A.}~\bibnamefont {Yazdani}},\
  }\bibfield  {title} {\bibinfo {title} {Tuning interactions between spins in a
  superconductor},\ }\bibfield  {journal} {\bibinfo  {journal} {Proceedings of
  the National Academy of Sciences}\ }\textbf {\bibinfo {volume} {118}},\ \href
  {https://doi.org/10.1073/pnas.2024837118} {10.1073/pnas.2024837118} (\bibinfo
  {year} {2021})\BibitemShut {NoStop}%
\bibitem [{\citenamefont {Kim}\ \emph {et~al.}(2015)\citenamefont {Kim},
  \citenamefont {Zhang}, \citenamefont {Rossi},\ and\ \citenamefont
  {Lutchyn}}]{Kim2015}%
  \BibitemOpen
  \bibfield  {author} {\bibinfo {author} {\bibfnamefont {Y.}~\bibnamefont
  {Kim}}, \bibinfo {author} {\bibfnamefont {J.}~\bibnamefont {Zhang}}, \bibinfo
  {author} {\bibfnamefont {E.}~\bibnamefont {Rossi}},\ and\ \bibinfo {author}
  {\bibfnamefont {R.~M.}\ \bibnamefont {Lutchyn}},\ }\bibfield  {title}
  {\bibinfo {title} {{Impurity-Induced Bound States in Superconductors with
  Spin-Orbit Coupling}},\ }\href
  {https://doi.org/10.1103/PhysRevLett.114.236804} {\bibfield  {journal}
  {\bibinfo  {journal} {Phys. Rev. Lett.}\ }\textbf {\bibinfo {volume} {114}},\
  \bibinfo {pages} {236804} (\bibinfo {year} {2015})}\BibitemShut {NoStop}%
\bibitem [{\citenamefont {Li}\ \emph {et~al.}(2014{\natexlab{b}})\citenamefont
  {Li}, \citenamefont {van~'t Erve}, \citenamefont {Robinson}, \citenamefont
  {Liu}, \citenamefont {Li},\ and\ \citenamefont {Jonker}}]{CHLi2014}%
  \BibitemOpen
  \bibfield  {author} {\bibinfo {author} {\bibfnamefont {C.~H.}\ \bibnamefont
  {Li}}, \bibinfo {author} {\bibfnamefont {O.~M.~J.}\ \bibnamefont {van~'t
  Erve}}, \bibinfo {author} {\bibfnamefont {J.~T.}\ \bibnamefont {Robinson}},
  \bibinfo {author} {\bibfnamefont {Y.}~\bibnamefont {Liu}}, \bibinfo {author}
  {\bibfnamefont {L.}~\bibnamefont {Li}},\ and\ \bibinfo {author}
  {\bibfnamefont {B.~T.}\ \bibnamefont {Jonker}},\ }\bibfield  {title}
  {\bibinfo {title} {{Electrical detection of charge-current-induced spin
  polarization due to spin-momentum locking in Bi$_2$Se$_3$}},\ }\href@noop {}
  {\bibfield  {journal} {\bibinfo  {journal} {Nature Nanotechnology}\ }\textbf
  {\bibinfo {volume} {9}},\ \bibinfo {pages} {218} (\bibinfo {year}
  {2014}{\natexlab{b}})}\BibitemShut {NoStop}%
\bibitem [{\citenamefont {Xie}\ \emph {et~al.}(2014)\citenamefont {Xie},
  \citenamefont {He}, \citenamefont {Chen}, \citenamefont {Feng}, \citenamefont
  {Yi}, \citenamefont {Liang}, \citenamefont {Zhao}, \citenamefont {Mou},
  \citenamefont {He}, \citenamefont {Peng}, \citenamefont {Liu}, \citenamefont
  {Liu}, \citenamefont {Liu}, \citenamefont {Dong}, \citenamefont {Yu},
  \citenamefont {Zhang}, \citenamefont {Zhang}, \citenamefont {Wang},
  \citenamefont {Zhang}, \citenamefont {Yang}, \citenamefont {Peng},
  \citenamefont {Wang}, \citenamefont {Chen}, \citenamefont {Xu},\ and\
  \citenamefont {Zhou}}]{ZXie2014}%
  \BibitemOpen
  \bibfield  {author} {\bibinfo {author} {\bibfnamefont {Z.}~\bibnamefont
  {Xie}}, \bibinfo {author} {\bibfnamefont {S.}~\bibnamefont {He}}, \bibinfo
  {author} {\bibfnamefont {C.}~\bibnamefont {Chen}}, \bibinfo {author}
  {\bibfnamefont {Y.}~\bibnamefont {Feng}}, \bibinfo {author} {\bibfnamefont
  {H.}~\bibnamefont {Yi}}, \bibinfo {author} {\bibfnamefont {A.}~\bibnamefont
  {Liang}}, \bibinfo {author} {\bibfnamefont {L.}~\bibnamefont {Zhao}},
  \bibinfo {author} {\bibfnamefont {D.}~\bibnamefont {Mou}}, \bibinfo {author}
  {\bibfnamefont {J.}~\bibnamefont {He}}, \bibinfo {author} {\bibfnamefont
  {Y.}~\bibnamefont {Peng}}, \bibinfo {author} {\bibfnamefont {X.}~\bibnamefont
  {Liu}}, \bibinfo {author} {\bibfnamefont {Y.}~\bibnamefont {Liu}}, \bibinfo
  {author} {\bibfnamefont {G.}~\bibnamefont {Liu}}, \bibinfo {author}
  {\bibfnamefont {X.}~\bibnamefont {Dong}}, \bibinfo {author} {\bibfnamefont
  {L.}~\bibnamefont {Yu}}, \bibinfo {author} {\bibfnamefont {J.}~\bibnamefont
  {Zhang}}, \bibinfo {author} {\bibfnamefont {S.}~\bibnamefont {Zhang}},
  \bibinfo {author} {\bibfnamefont {Z.}~\bibnamefont {Wang}}, \bibinfo {author}
  {\bibfnamefont {F.}~\bibnamefont {Zhang}}, \bibinfo {author} {\bibfnamefont
  {F.}~\bibnamefont {Yang}}, \bibinfo {author} {\bibfnamefont {Q.}~\bibnamefont
  {Peng}}, \bibinfo {author} {\bibfnamefont {X.}~\bibnamefont {Wang}}, \bibinfo
  {author} {\bibfnamefont {C.}~\bibnamefont {Chen}}, \bibinfo {author}
  {\bibfnamefont {Z.}~\bibnamefont {Xu}},\ and\ \bibinfo {author}
  {\bibfnamefont {X.~J.}\ \bibnamefont {Zhou}},\ }\bibfield  {title} {\bibinfo
  {title} {Orbital-selective spin texture and its manipulation in a topological
  insulator},\ }\href@noop {} {\bibfield  {journal} {\bibinfo  {journal}
  {Nature Communications}\ }\textbf {\bibinfo {volume} {5}},\ \bibinfo {pages}
  {3382} (\bibinfo {year} {2014})}\BibitemShut {NoStop}%
\end{thebibliography}
%


\end{document}